%% file: ms_141124_vbgb.tex
\documentclass[12pt]{emulateapj}

\usepackage{txfonts}
\def\ltsima{$\; \buildrel < \over \sim \;$}
\def\simlt{\lower.5ex\hbox{\ltsima}}
\def\gtsima{$\; \buildrel > \over \sim \;$}
\def\simgt{\lower.5ex\hbox{\gtsima}}
\def\kms{{\rm\,km\,s^{-1}}}
\def\mas{{\rm\,mas}}

\def\kpc{{\rm\,kpc}}

\def\msun{{\rm\,M_\odot}}
\def\lsun{{\rm\,L_\odot}}

\def\pc{{\rm\,pc}}

\def\hst{{HST\/}}
\def\bmv{\hbox{\it B--V\/}}
\def\bmr{\hbox{\it B--R\/}}
\def\bmi{\hbox{\it B--I\/}}
\def\vmi{\hbox{\it V--I\/}}

\def\jmh{\hbox{\it J--H\/}}
\def\jmk{\hbox{\it J--K\/}}
\def\hmk{\hbox{\it H--K\/}}
\def\bmk{\hbox{\it B--K\/}}
\def\vmr{\hbox{\it V--R\/}}
\def\rmi{\hbox{\it R--I\/}}
\def\bmj{\hbox{\it B--J\/}}
\def\vmj{\hbox{\it V--J\/}}
\def\rmj{\hbox{\it R--J\/}}
\def\imj{\hbox{\it I--J\/}}
\def\bmh{\hbox{\it B--H\/}}
\def\vmh{\hbox{\it V--H\/}}
\def\rmh{\hbox{\it R--H\/}}
\def\imh{\hbox{\it I--H\/}}
\def\vmk{\hbox{\it V--K\/}}
\def\rmk{\hbox{\it R--K\/}}
\def\imk{\hbox{\it I--K\/}}

\shorttitle{}
\shortauthors{Braga et al.}

\begin{document}
\title{On the distance of the globular cluster M4 (NGC~6121) using RR Lyrae stars: 
I. optical and near-infrared Period-Luminosity and Period-Wesenheit relations}

\author{V.~F.~Braga\altaffilmark{1,2}, 
M.~Dall'Ora\altaffilmark{3},
G.~Bono\altaffilmark{1,2},
P.~B.~Stetson\altaffilmark{4},
I.~Ferraro\altaffilmark{2}, 
G.~Iannicola\altaffilmark{2}, 
M.~Marengo\altaffilmark{5}, 
J.~Neeley\altaffilmark{5},
S.~E.~Persson\altaffilmark{6},
R.~Buonanno\altaffilmark{1,7},
G.~Coppola\altaffilmark{3},
W.~Freedman\altaffilmark{6},
B.~F.~Madore\altaffilmark{6},
M.~Marconi\altaffilmark{3},
N.~Matsunaga\altaffilmark{8},
A.~Monson\altaffilmark{6},
J.~Rich\altaffilmark{6},
V.~Scowcroft\altaffilmark{6},
M.~Seibert\altaffilmark{6}}
   
\altaffiltext{1}{Department of Physics, Universit\`a di Roma Tor Vergata, via della Ricerca Scientifica 1, 00133 Roma, Italy}
\altaffiltext{2}{INAF-Osservatorio Astronomico di Roma, via Frascati 33, 00040 Monte Porzio Catone, Italy}
\altaffiltext{3}{INAF-Osservatorio Astronomico di Capodimonte, Salita Moiariello 16, 80131 Napoli, Italy}
\altaffiltext{4}{NRC-Herzberg, Dominion Astrophysical Observatory, 5071 West Saanich Road, Victoria BC V9E 2E7, Canada}
\altaffiltext{5}{Department of Physics and Astronomy, Iowa State University, Ames, IA 50011, USA}
\altaffiltext{6}{Carnegie Observatories, 813 Santa Barbara Street, Pasadena, CA 91101, USA}
\altaffiltext{7}{INAF-Osservatorio Astronomico di Teramo, Via Mentore Maggini snc, Loc. Collurania, 64100 Teramo, Italy}
\altaffiltext{8}{Kiso Observatory, Institute of Astronomy, School of Science, The University of Tokyo 10762-30, Mitake, Kiso-machi, Kiso-gun, 3 Nagano 97-0101, Japan}

\date{\centering drafted \today\ / Received / Accepted }

\begin{abstract}
We present new distance determinations to the nearby globular
M4 (NGC~6121) based on accurate optical and Near Infrared (NIR)
mean magnitudes for fundamental (FU) and first overtone (FO) RR Lyrae variables (RRLs), and
new empirical optical and NIR Period-Luminosity (PL)
and Period-Wesenheit (PW) relations. We have found that optical-NIR and NIR PL and
PW relations are affected by smaller standard deviations than optical relations. 
The difference is the consequence of a steady decrease in the
intrinsic spread of cluster RRL apparent magnitudes at fixed period
as longer wavelengths are considered.
The weighted mean visual apparent magnitude of 44 cluster RRLs is
$\left<V\right>=13.329\pm0.001$ (standard error of the mean) $\pm$0.177
(weighted standard deviation) mag.
Distances were estimated using RR Lyr itself to fix the zero-point
of the empirical PL and PW relations. Using the entire sample
(FU$+$FO) we found weighted mean true distance moduli of 11.35$\pm$0.03$\pm$0.05 mag and
11.32$\pm$0.02$\pm$0.07 mag. Distances were also evaluated using predicted
metallicity dependent PLZ and PWZ relations.
We found weighted mean true distance moduli of 11.283$\pm$0.010$\pm$0.018
mag (NIR PLZ) and 11.272$\pm$0.005$\pm$0.019 mag (optical--NIR and NIR PWZ).
The above weighted mean true distance moduli agree within 1$\sigma$.
The same result is found from distances based on PWZ relations in which the
color index is independent of the adopted magnitude (11.272$\pm$0.004$\pm$0.013 mag).
These distances agree quite well with the geometric distance provided by
\citep{kaluzny2013} based on three eclipsing binaries. The available evidence
indicates that this approach can provide distances to globulars hosting
RRLs with a precision better than 2--3\%.
\end{abstract}

\maketitle

%%%%%%%%%%%%%%%%%%%%%%%%%%%%%%%%%%%%%%%%%%%%%%%%%%%%%%%%%%%%%%%%%%%%%%%%%%%%%%%%
\section{Introduction} \label{chapt_intro}

Globular clusters (GCs) have played a crucial role in modern astrophysics. 
They are fundamental laboratories not only for stellar evolution
\citep{denissenkov2003,pietrinferni2006,dotter2007,vandenberg2012,pietrinferni2013} 
and stellar dynamics \citep{kouwenhoven2010}, but also for constraining models 
of Galaxy formation and evolution 
\citep{zoccali2000,marinfranch2009,valenti2011,Leaman2013,johnson2013}
and primordial abundances \citep{Zoccali2003,salaris2004,troisi2011}

It is not surprising that the astronomical community has carried out 
an enormous theoretical and observational effort to properly constrain 
their structural parameters \citep{casettidinescu2013,dicecco2013} 
and intrinsic properties \citep{gratton2004,bono2010,milone2013}

Dating back to the seminal investigations by \citet{zinn1980} and \citet{zinnwest1984} 
and to the more recent analysis of iron \citep{kraft2003,carr09} 
and $\alpha$-element abundances we have solid estimates of the metallicity 
scale in Galactic globulars. The same is true for the abundances of s- 
and r-process elements \citep{roederer2011,lardo2013}
and of lithium \citep{spite2012}.

During the last few years we have also acquired a wealth of new information  
on the kinematic properties of halo and bulge Galactic globulars 
\citep{casettidinescu2007,vieira2007,casettidinescu2013,poleski2013}. 
Detailed numerical simulations have also been provided for the survival 
rate of globulars after multiple bulge and disk crossings \citep{capuzzo2008}.  

The estimation of both absolute and relative ages of Galactic globulars has been 
at the crossroads of several detailed investigations 
\citep{buonanno1998,stetson1999,Zoccali2003,richer2004,deangeli2005,richer2013}.
% Referee 2
The recent survey based on photometry with the Advanced Camera for Surveys on board the Hubble Space Telescope (\hst)
has been applied to large samples of 
Galactic globulars. They range from homogeneous relative ages for 
nine GCs by \citet{sarajedini2007} to 64 GCs by \citet{marinfranch2009}, 
to six GCs by \citet{dotter2011} and to 55 GCs of \citet{vandenberg2013}.

% Referee 1_i
The scenario outlined above indicates that we are dealing with precise 
and homogeneous investigations concerning age and metallicity distributions and 
the kinematics of Galactic globulars. However, we still lack a 
homogeneous distance scale for GCs. The reasons are manifold: 

i) The primary distance indicators adopted to estimate absolute distances of 
GCs can only be applied to subsamples. The tip of the red giant branch can 
be applied reliably only to very massive GCs, namely $\omega$~Cen and 47 Tuc. The white dwarf 
cooling sequence has only been applied to nearby GCs \citep{zoccali2001,richer2013}. 
Main sequence fitting has only been applied to GCs 
with iron abundances bracketed by nearby dwarf stars with accurate 
trigonometric parallaxes \citep{gratton2003,bond2013}.     
The use of kinematic distances has also been applied only to nearby GCs \citep{peterson1995,layden2005}. 
Distances from eclipsing binaries are very precise 
and promising, but they have only been measured for a limited sample \citep{thompson2010,kaluzny2013}.
% Referee 3_i
The use of the predicted Zero-Age-Horizontal-Branch (ZAHB) luminosity appears as a very 
promising approach \citep{vandenberg2013}. However, uncertainties in the input physics 
(electron conductive opacities, \citet{cassisi2007}) and in the mass loss rate during the
Red Giant Branch (RGB) and Horizontal-Branch (HB) evolutionary phases (\citet{salaris2012}) 
affect the predicted luminosity of HB stellar structures.

The luminosity of the AGB bump has several advantages, but its application is once again 
limited to massive GCs \citep{pulone1992,salaris2013}. The Red Giant Branch (RGB) bump 
is also an interesting    
distance indicator, but predicted luminosities are at odds with observed luminosities 
and we still lack an accurate empirical calibration \citep{ferraro1999}. 
 
% Referee 5_iv  
ii) The Leavitt relation of type II Cepheids and MIRAS has also been applied 
to a limited number of GCs \citep{feast2000, matsunaga2009}. The Leavitt 
relation is a Period-Magnitude relation, but we will refer to it as a 
Period-Luminosity (PL) relation to point out the difference with the 
Period-Wesenheit (PW) relation.
The $M_V$ vs iron relation for RR Lyraes (RRLs) has been applied to several GCs, but their 
distances are affected by evolutionary effects and by a possible nonlinearity 
of the relation \citep{caputo2000}.
The SX Phoenicis stars have also been used to estimate the distances of a few 
GCs \citep{gilliland1998,kaluzny2009,mcnamara2011,cohen2012}, 
but the physical mechanisms driving their formation and identification of their pulsation mode 
are still controversial \citep[][ and references therein]{fiorentino2013}.   

iii) Several of the above methods are affected by uncertainties in the 
cluster reddening. This problem becomes even more severe for GCs affected 
by differential reddening. We still lack a reddening scale based on a single 
diagnostic that can be used for both halo and bulge GCs.

% Referee 3_ii
iv) Photometry and spectroscopy of cluster stars located in the innermost regions is often 
a difficult observational problem due to crowding. Recent empirical evidence indicates that 
cluster RRLs located in the central cluster regions might be contaminated by neighboring 
stars \citep{majaess2012a,majaess2012b}. It is worth mentioning that the central density of 
M4 ($\log$\,$\rho_V$=3.64 $\lsun$/\pc$^3$) is one to two orders of magnitude smaller than GCs 
with high central densities \citep[$\log$\,$\rho_V$=4.6--5.6 $\lsun$/\pc$^3$; ][]{harris96}
Additionally, the half-light radius of M4 is among the largest, at
4.33 arcmin \citep[][]{harris96}.  For these reasons M4 is not nearly as severely affected 
by crowding problems as the bulk of Galactic globulars.

% Referee 1_ii
The theoretical and empirical scenario concerning absolute and relative 
distances to Galactic globulars \citep{bono08a} is far from being    
satisfactory. Precise distances based on geometrical methods are 
limited to only a few nearby clusters. Moreover, the different standard candles 
are still affected by systematics that need to be constrained by independent 
and precise diagnostics.

In this investigation we provide a new estimate of the true
distance modulus of M4 from new optical ({\it UBVRI\/}) and Near Infrared (NIR, {\it JHK\/})
photometry for RRLs in the cluster \citep{stetson2014}. For
this purpose we use optical and NIR Period-Luminosity-Metallicity (PLZ) and
Period-Wesenheit-Metallicity (PWZ) relations; the latter provide distances that are
corrected for reddening, assuming that the reddening law is known.

The structure of the paper is as follows. In \S 2 we discuss 
recent distance determinations to M4, while in \S 3 we present the 
optical and NIR datasets used in this 
investigation. Then \S 4 deals with the observed optical and 
NIR PL relations; moreover, here
we also compare to similar results available in the literature. 
Empirical optical, optical-NIR and NIR PWZ
relations are discussed in \S 5. In \S 6 we present new theoretical 
optical and NIR PLZ and PWZ relations. The true 
distance moduli based on the current optical and NIR photometry 
are discussed in \S 7. Finally, in \S 8 we summarize the results of 
this investigation and briefly outline the anticipated future development of 
the project. 

%_____________________________________________________________________
\section{Distance evaluations to the GC M4} \label{chap_intro_m4_distance}

The distance to M4 has been estimated using several primary
distance indicators, since it is the closest GC. \citet{peterson1995} obtained a geometric
distance by comparing the radial-velocity and proper-motion
dispersions, finding a true distance modulus of 11.18$\pm$0.18
mag. The M4 distance was also estimated by \citet{liujanes1990b}, who
applied the infrared surface-brightness technique---a variant of
the Baade-Wesselink method---to four cluster RRLs (V2, V15,
V32, V33); they found a true distance modulus of 11.19$\pm$0.01
mag. Note that the stated error is only the standard error of the
mean distance for the four RRLs, and does not take account of
possible systematic uncertainties such as the $p$-factor that has
been adopted, i.e., the parameter that transforms the observed
radial velocity into a pulsation velocity \citep{nard2013}. 
The current uncertainties in the $p$-factor imply {\it systematic\/} uncertainties 
in individual RRLs distances of the order of 10\% \citep{marconi2005}.

A similar distance to M4 was also obtained by \citet{longmore1990} in their
seminal investigation of the $K$-band PL relation for
cluster RRLs.  Applying a new calibration of the $K$-band PL
relation to NIR photometry of 26 RRLs they found a true
distance modulus of 11.28$\pm$0.06 mag for an assumed $E(\bmv)$ of
0.37$\,$mag. A similar approach was also adopted by  \citet{bono2003},
but they employed a $K$-band PLZ relation based on nonlinear pulsation models. They used the four RRLs
with accurate $K$-band light curves and individual
reddening estimates \citep{liujanes1990b} and, assuming an iron
content of [Fe/H]=--1.30 (see their Table~6), they found a true
distance modulus of 11.37$\pm$0.08 mag.     

The distance to M4 has also been estimated by \citet{hendricks2012} (henceforth H12)
from the HB luminosity level, and they found a true distance modulus of
11.28$\pm$0.06 (random error) and a mean reddening
$E(\bmv)=0.37\pm0.01$ mag. Note that H12 also considered 
uncertainties in the extinction parameter---$R_V$---adopted
in the empirical reddening law \citep{cardelli} to constrain the
selective absorption coefficients (see their Table~5).  They concluded that
a value of $R_V \sim 3.6$ was preferable to the canonical value of $\sim 3.1$, presumably related to the
$\rho$~Oph star-forming cloud lying in front of the cluster.

Main-sequence fitting to field subdwarfs was adopted by
\citet{richer1997} and by \citet{hansen2004} to estimate the 
distance; they found a true distance modulus 11.18$\pm$0.18
mag for an assumed reddening of $E(\bmv)=0.35\pm0.01$ mag and a
ratio of total to selective extinction $R_V$=3.8. More recently,
\citet{kaluzny2013} used three detached eclipsing double-lined
binary members of M4 and the reddening law found by H12 (see
Table~\ref{tab:table_dis}) to provide a true distance 
modulus of 11.30$\pm$0.05 mag.

% Referee 1_iii
The distance determinations discussed in this section suggest that estimates of the 
absolute distance to the closest GC range from 1.72$\pm$0.14 \kpc~\citep{peterson1995} 
to $\sim$1.98 \kpc ~\citep{bedin2009}. They agree within 1$\sigma$, but the full size of the confidence interval 
is of the order of 15\% (see distance determinations listed in 
Table~\ref{tab:table_dis}).

%_______________________________________________________________________________
\section{Optical and near-infrared data sets}\label{chapt_data}

The reader interested in a detailed discussion of 
the different optical and NIR data sets and the approach adopted to perform 
the photometry is referred to \citep{stetson2014}. 
The optical photometry was transformed into the Johnson ({\it UBV\/}), 
Kron/Cousins ({\it RI\/}) photometric system \citep{stetson2000,stetson2005}.  
The NIR photometry was transformed into the 2MASS {\it JH}$K_s$ photometric system 
\citep{skrutskie06}.

In the following we neglect the available $U$-band photometry because of the relatively poor time 
sampling and the limited accuracy of individual measurements.
The optical light curves are characterized by good time sampling 
and the number of measurements ranges from 900 to 1100 in the $B$ band, 
from 1400 to 1500 in the $V$ band, from 1580 to 1800 in the $R$ band,
and from ten to 60 in the $I$ band. The NIR light curves have more limited 
coverage and the number of measurements ranges from five to 55 in the 
$J$ band, from one to nine in the $H$ band, and from two to 40 in the 
$K$ band. 

The mean optical and NIR magnitudes were 
evaluated as intensity means and then transformed into magnitude. 
The phasing of the light curves was performed with the new
period estimates provided by \citet{stetson2014}. The mean magnitudes 
in the bands with good time sampling  ({\it BVRJ\/}) were estimated 
from a fit with a spline under tension. The individual mean 
magnitudes were estimated by equally sampling the analytical fit. 
The mean magnitudes in the $I$ and $K$ bands were estimated 
using the light curve templates provided by \citet{dicriscienzo2011} 
and \citet{jones96}.  
To apply the templates we adopted the epochs of maxima and the optical 
amplitudes provided by \citet{stetson2014}. For two variables not 
covered by our optical photometry, we adopted epochs of maxima and 
amplitudes available in the literature. The reader interested in 
a more detailed discussion concerning the amplitude ratio between 
optical and NIR magnitudes is referred to \citet{stetson2014}. 

The number of candidate cluster RRLs is currently 44 
(31 fundamental = ``FU'' pulsators, and 13 first-overtone = ``FO'' pulsators) 
and their periods range from 0.2275 to 0.6270 days 
plus a single long-period FU RRLs with P=0.8555 days. The presence 
of such a long-period variable is consistent with the tail in the 
period distribution of $\omega$~Cen RRLs found by \citet{marconi2011}.
On the other hand, \citet{andrievsky2010} suggested that at
least some field long-period RRLs, such as KP Cyg, appear to be
metal-rich plus C- and N-enhanced. Therefore, they suggested that these
objects could be short period BL Her stars, defining a new class of
variable stars, instead of long-period RRL. The extended spectroscopic analysis
of both evolved and main sequence stars performed by \citet{malavolta2014}
does not support the presence of a spread in metal abundance in M4. However,
we still lack detailed information concerning CNO abundances among cluster HB stars.

%_______________________________________________________________________________
\section{Observed Optical and NIR Period-Luminosity relations}\label{chapt_pl}

The current empirical and theoretical evidence indicates that the RRLs do obey 
a PL relation. The key feature is that the slope is negative 
for wavelengths longer than the $V$ band, while it is positive at shorter 
wavelengths. In the $V$ band the slope attains a negligible value. This is the 
main reason why the $M_V^{RR}$ vs [Fe/H] relation was so popular in the last 
century to estimate the distance of both cluster and field RRLs. 
%
% Referee 8 
Plain physics arguments suggest that the occurrence of well defined NIR PL relations for RRLs
is due to a significant change in the NIR bolometric corrections when moving from 
the blue (short periods) to the red (long period) edge of the instability strip. This change 
is vanishing in the $V$ band and becomes of the order of 1.5 magnitudes in the $K$ band 
\citep[see Fig.~1 in ][]{bono2003a}. This working hypothesis was also supported in an 
independent theoretical investigation by \citet{catelan2004}. Further evidence for a lack 
of PL relations in $B$ and $V$ bands was brought forward by \citet{benko2006} using 
accurate photometry for more than 220 RRLs in M3 (see their Fig.~8).
The NIR PL relations of RRLs became quite popular as distance 
indicators soon after their empirical determination 
\citep{longmore1986,jones1988,longmore1990,liujanes1990,liujanes1990b,carney1992}.
A new spin was then provided by theoretical pulsation and evolutionary 
predictions \citep{bono2001,bono2003,catelan2004,cassisi2004}.
%Referee 4_i
The observational scenario was also significantly improved by the use of the new NIR arrays 
\citep{dallora2004,delprincipe2005}, the 2MASS photometry \citep{sollima2006}, and the first 
accurate trigonometric parallax for RR Lyr itself \citep{vanaltena1995,perryman1997,benedict2002}.
The trigonometric parallax of RR Lyr was also adopted to fix the 
zero-point of both the theoretical \citep{bono2002} and empirical $K$--band PL relations 
\citep{sollima2006,sollima2008}.

Pros and cons of optical-NIR PL relations and of the $M_V^{RR}$ vs [Fe/H]  
relations have been widely discussed in the literature. In passing, we mention 
that the optical, NIR and mid-infrared (MIR) PL relations for RRLs 
appear to be linear 
over the entire period range \citep{bono2001,bono2003,catelan2004,madore2013}. 
The above diagnostics are prone to uncertainties in the reddening corrections 
and in the adopted reddening law. Obviously, the problem becomes less and less 
severe when moving from the optical to the NIR and MIR bands. The 
impact when compared with the $V$ band is ten times smaller in the $K$ band and 
more than 20 times smaller in the $3.6\,\mu$m band.

The key advantage in dealing with the RRLs in M4 is that 
quantities necessary for calculating the distance moduli, such as the mean 
reddening ($E(\bmv)=0.37\pm0.01$ mag), the ratio of total to selective absorption 
($R_V$=3.62 $\pm$ 0.07 mag) and the overall reddening law have been recently 
provided by H12 (see their Table~5).

On the basis of the current mean magnitudes we estimated optical ({\it RI\/}) and 
NIR ({\it JHK}) PL relations. We decided to provide independent PL relations 
for FO and FU pulsators. The reasons are threefold. i) Empirical and theoretical 
evidence indicates that the width in temperature of the region in which FO 
variables are pulsationally stable is roughly a factor of two narrower than 
the region in which FU variables are pulsationally stable.    
This means that the standard deviations of FO PL relations are, at fixed 
photometric precision, intrinsically smaller than for FU PL relations. 
ii) Smaller standard deviations imply more accurate relative and 
absolute distances. iii) The light curves of FO pulsators are more 
nearly sinusoidal and show pulsation amplitudes that are on average 
from 2 to 3 times smaller than FU variables. This means that a more limited 
number of measurements can provide accurate mean magnitudes.
The main drawback is that FO variables are typically $\sim$0.5 mag 
fainter than FU variables in the longer-wavelength bands.

However, the number of FO variables in M4 is modest 
and to improve the precision of the empirical PL relations we also derived 
PL relations from the entire sample of FU and FO variables. The global 
PL relations were evaluated by fundamentalizing FO periods according to 
the relation $\log P_F$=$\log P_{FO}$+0.127. This approach to fundamentalizing
the period of first-overtone variables relies on the assumption that the period ratio 
of double-mode RRLs attains a constant value of the order of 0.746. 
The above assumption was supported by former theoretical and empirical 
evidence \citep{iben1971,rood1973,cox1983}. However, recent 
findings indicate that the double-mode field and cluster RRLs do cover 
a significant range in period ratios ($\sim$0.735--0.750, Coppola et al. 2014, 
in preparation). The same outcome results from nonlinear pulsation predictions 
(Marconi et al. 2014, in preparation). We plan to address this issue in a future 
paper.  
The zero-points, the slopes, and
their errors and standard deviations are listed in Table~\ref{tab:table_pl}. 
The data given in this Table support the above contention that the 
standard deviations of FO PL relations are smaller than FU PL relations, 
and these are in turn smaller than FU$+$FO PL relations (see vertical error bars). 
There is also evidence that the zero-points and the slopes of both optical 
and NIR PL relations agree within one $\sigma$. This finding might be 
affected by the limited sample of FO variables in M4. We plan to address 
this issue in a future paper in which we will deal with larger samples of 
cluster variables (Braga et al. in preparation).  

\begin{figure*}[tbp]
\centering
\includegraphics[width=13cm,height=13cm]{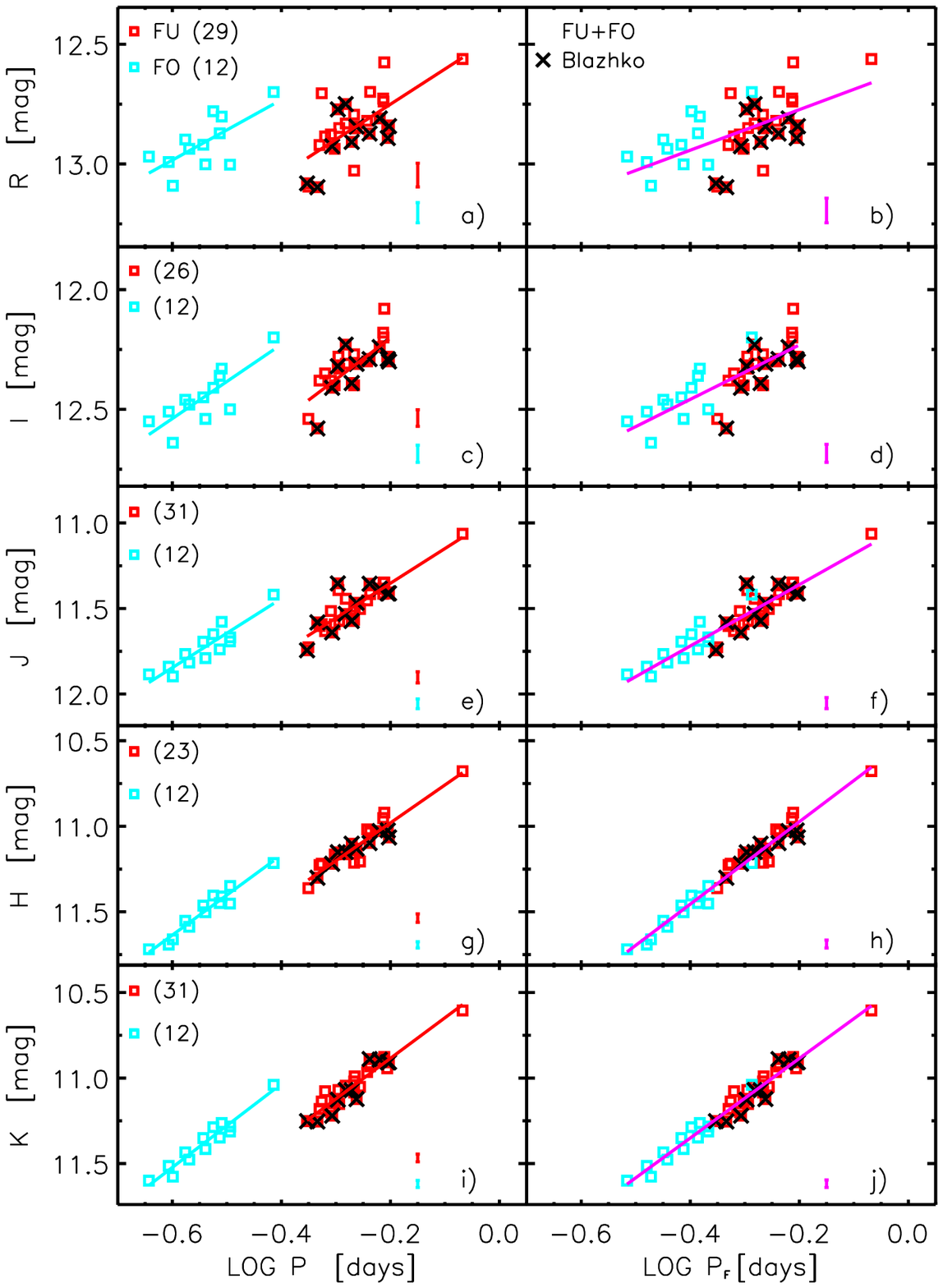}
\vspace*{0.5truecm}
\caption{From top to bottom observed optical and NIR Period--Luminosity 
(PL) relations for RR Lyrae in M4.    
Panel a): $R$--band Period--Luminosity (PL) relation. Fundamental (FU) and first overtone (FO) 
pulsators are marked with red diamonds and cyan squares, respectively. 
The black crosses display candidate Blazkho RR Lyrae. 
The cyan and the red lines display the linear fits, while the vertical 
bars show the standard deviations, $\sigma$, of the fits. The number of 
variables adopted in the fits are also labeled.
Panel b): Same as panel a), but for FU and FO RR Lyrae. The periods of 
FO variables were fundamentalized using the relation: 
$\log  P_{FU}$= $\log  P_{FO}$+0.127.
Panels c) and d): Same as panels a) and b), but for the $I$--band PL relation. 
Panels e) and f): Same as panels a) and b), but for the $J$--band PL relation. 
Panels g) and h): Same as panels a) and b), but for the $H$--band PL relation. 
Panels i) and j): Same as panels a) and b), but for the $K$--band PL relation.}  
\label{fig1}
\end{figure*}

The data plotted in Fig.~\ref{fig1} show that the intrinsic dispersion  
of the PL relations decreases steadily when moving from the optical to the 
NIR bands. The standard deviation in the $R$ band is a factor of two larger than 
in the $K$ band. The reasons for the difference were mentioned above. In passing,
we also note that the slope of the $K$-band PL relation is a factor of three steeper 
than the $R$-band PL relation. This means that the use of a PL instead of a 
Period-Luminosity-Color (PLC) relation---i.e., neglecting the 
width in temperature of the instability strip---becomes more valid when 
moving from the optical to the NIR \citep{bono2010b,coppola2011} 
and MIR \citep{madore2013} bands.   

Finally, we mention that candidate Blazhko RRLs (black crosses) seem 
to follow, within the errors, PL relations similar to singly periodic 
FU variables.

%_______________________________________________________________________________
\section{Observed optical, optical--NIR and NIR Period--Wesenheit relations}\label{chapt_plw}

The key advantages in using PW relations in estimating individual 
distances are several. i) They are independent of reddening uncertainties
and of differential reddening provided the form of the reddening law is known \citep{vandenbergh1975,madore1982}. 
ii) They mimic a Period-Luminosity-Color relation so they can 
provide, in contrast with the Period-Luminosity relation, individual distances 
\citep{bono08b,inno2013}. They are also affected by two drawbacks. 
i) They require accurate mean magnitudes in a minimum of two photometric bands, and this limitation becomes
more severe in dealing with optical-NIR photometry; ii) they 
rely on the assumption that the reddening law is known. It is well understood that
this working hypothesis is not always valid in low-latitude Galactic regions. 
This limitation does not apply to M4, however, since H12 derived a reddening law specific to M4 
by considering both optical and NIR photometry.
Our current Wesenheit magnitudes have been estimated using the absorption 
coefficient ratios listed in Table~5 of H12. 

\begin{figure*}[tbp]
\centering
\includegraphics[width=14cm,height=15cm]{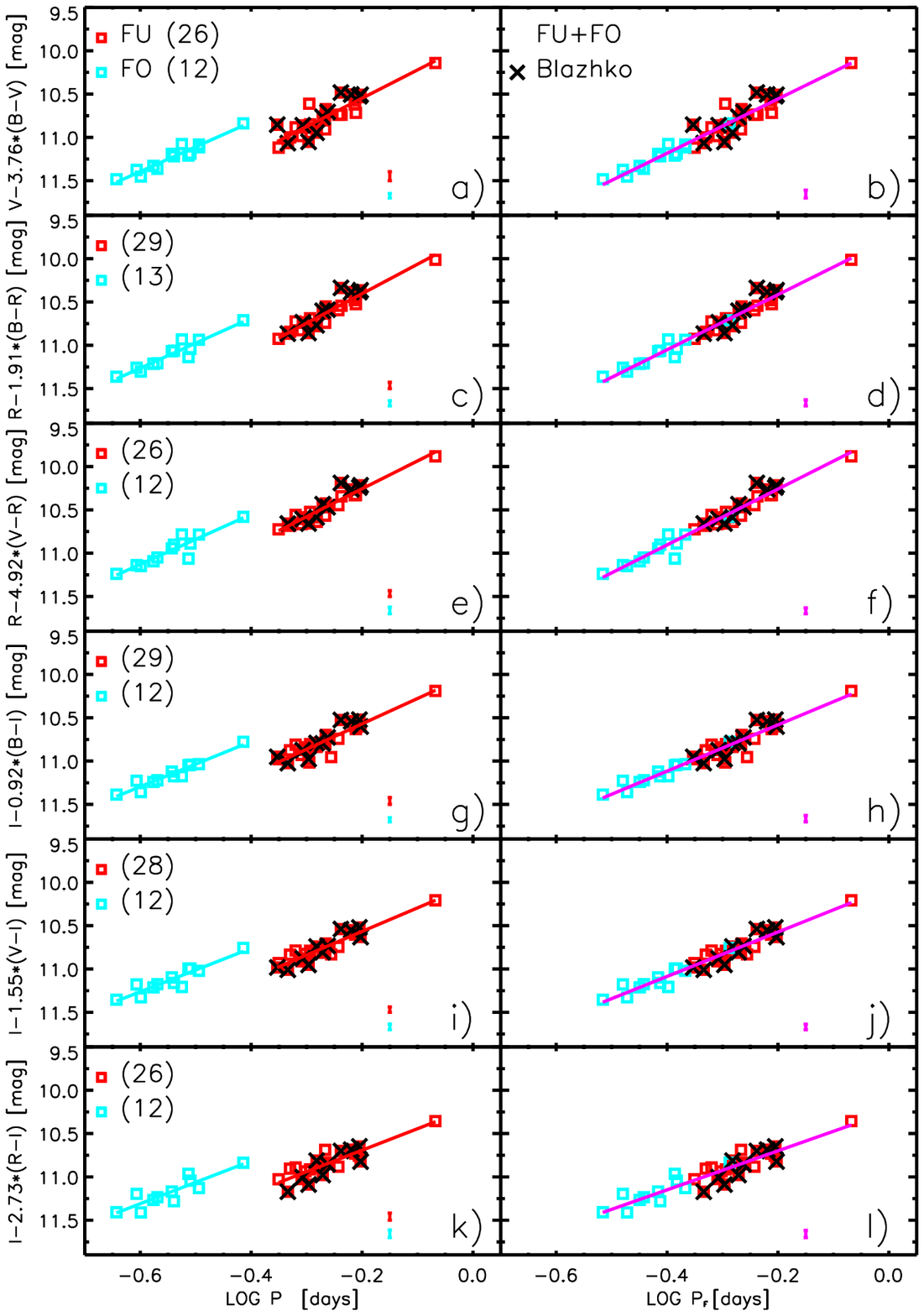}
\vspace*{0.5truecm} 
\caption{From top to bottom observed optical Period--Wesenheit (PW) 
relations for RR Lyrae in M4. Symbols and lines are the same as in Fig.~1. 
Panel a): PW($V$,\bmv) relation for FU and FO 
pulsators. 
Panel b):  Same as Panel a), but for the entire sample of RR Lyrae. The 
periods of FO variables were fundamentalized using the relation: 
$\log  P_{FU}$= $\log  P_{FO}$+0.127.    
Panels c) and d): Same as panels a) and b), but for the PW($R$,\bmr) relation. 
Panels e) and f): Same as panels a) and b), but for the PW($R$,\vmr) relation. 
Panels g) and h): Same as panels a) and b), but for the PW($I$,\bmi) relation. 
Panels i) and j): Same as panels a) and b), but for the PW($I$,\vmi) relation. 
Panels k) and l): Same as panels a) and b), but for the PW($I$,\rmi) relation. 
}\label{fig2}
\end{figure*}

Figure \ref{fig2} displays the six optical PW relations for FU and FO pulsators.
These show several distinctive features.

i) The intrinsic dispersion is, at fixed period, smaller than in the
optical PL relations. This difference is expected because we already 
mentioned that the PW relations mimic a PLC relation.  
Moreover, the dispersion of the individual data points decreases in 
Wesenheit magnitudes based on photometric bands that have a large difference in central wavelengths.  
This is caused by the fact that the coefficients of the color term in the 
Wesenheit magnitudes may attain values smaller than unity. This means that the 
$W$($I$,\bmi) index is less prone to uncertainties affecting the mean 
color than the $W$($I$,\rmi) index (0.92 vs 2.73). Moreover, the increased 
difference in central wavelength means also an increased sensitivity to 
the mean effective temperature of the variable.  

ii) The global PW relations including both FU and FO pulsators 
are characterized by smaller intrinsic dispersions and by smaller errors
in both the zero-point and the slope (see Table \ref{tab:table_plw_opt}) 
when compared with FU and FO individual PW relations).  
This difference is caused by the increase in sample size and, in particular, 
by the larger range in period covered by FU plus FO variables. However, 
the global PW relations might be affected by the assumption that the 
difference between FU and FO PW relations is only a difference in 
the zero point. The slopes of the global PW relations do attain 
values that are intermediate between the slopes of FU and FO PW 
relations. However, the current sample does not allow us to 
constrain this effect quantitatively.

iii) The slopes of the PW relations listed in columns 3, 6 and 9  
of Table \ref{tab:table_plw_opt} become steeper, as expected, when moving 
from Wesenheit magnitudes based on $V$ and $R$ magnitudes to those based 
on $I$ magnitudes \citep{catelan2004,bono2010b,madore2013}. 

iv) The candidate Blazhko variables display, within the errors,  
similar slopes to those of the canonical FU variables. This further supports the 
inference that the mean magnitudes and colors of the Blazhko variables are  
minimally affected by the secondary modulations, if the primary modulation 
is properly covered in the two adopted bands. 

\begin{figure*}[tbp]
\centering
\includegraphics[width=13cm,height=14cm]{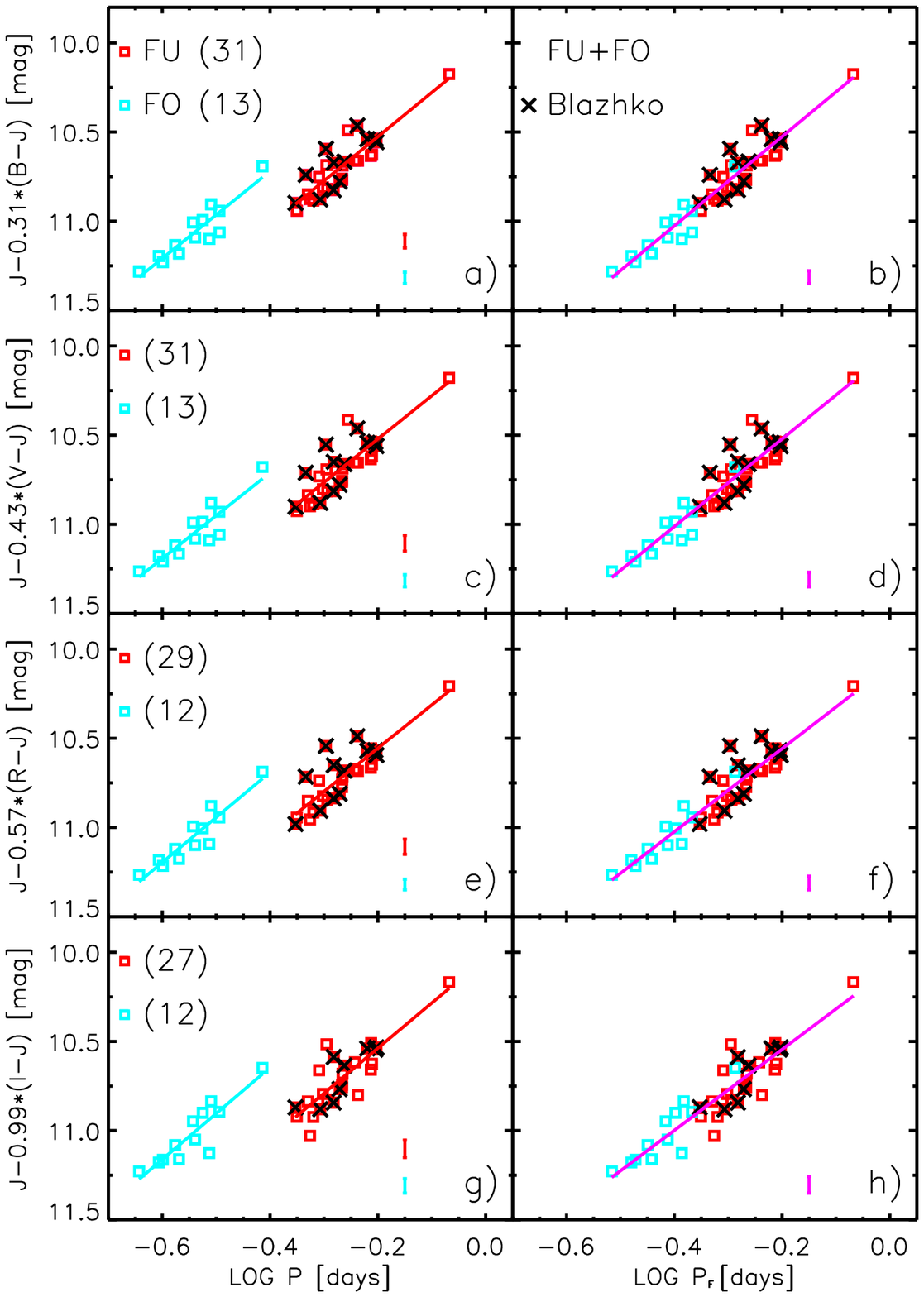}
\vspace*{0.5truecm} 
\caption{From top to bottom observed optical--NIR PW relations 
for RR Lyrae in M4. Symbols and lines are the same as in Fig.~1. 
Panel a): PW($J$,\bmj) relation for FU and FO
pulsators. 
Panel b): Same as panel a), but for the entire sample of RR Lyrae. The 
periods of FO variables were fundamentalized using the relation: 
$\log  P_{FU}$= $\log  P_{FO}$+0.127.    
Panels c) and d): Same as panels a) and b), but for the PW($J$,\vmj) relation. 
Panels e) and f): Same as panels a) and b), but for the PW($J$,\rmj) relation. 
Panels g) and h): Same as panels a) and b), but for the PW($J$,\imj) relation. 
}\label{fig3}
\end{figure*}

\begin{figure*}[tbp]
\centering
\includegraphics[width=13cm,height=14cm]{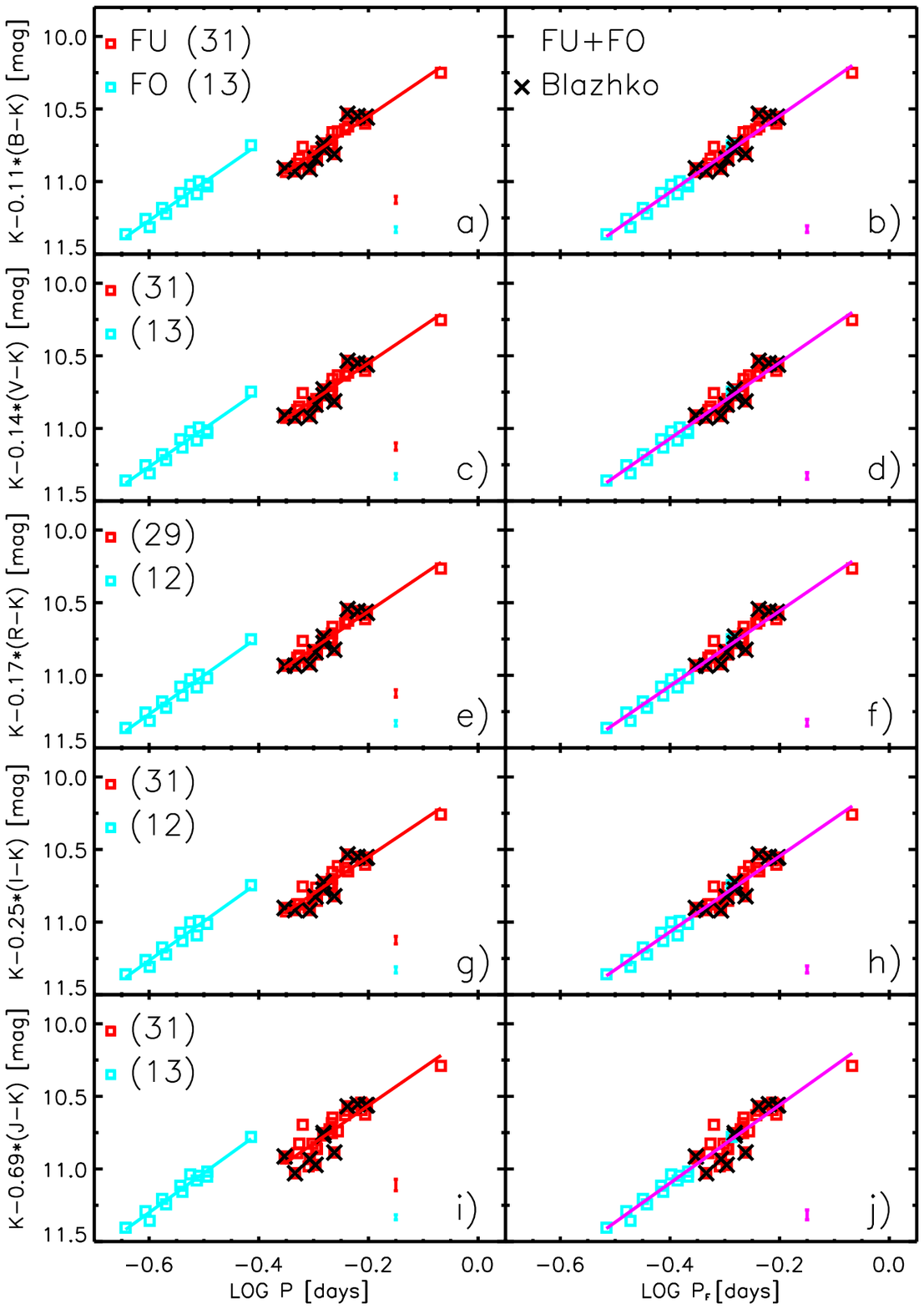}
\vspace*{0.5truecm} 
\caption{From top to bottom observed optical--NIR and NIR PW relations 
for RR Lyrae in M4. Symbols and lines are the same as in Fig.~1. 
Panel a): PW($K$,\bmk) relation for FU and FO pulsators. 
Panel b): Same as panel a), but for the entire sample of RR Lyrae. The 
periods of FO variables were fundamentalized using the relation: 
$\log  P_{FU}$= $\log  P_{FO}$+0.127.    
Panels c) and d): Same as panels a) and b), but for the PW($K$,\vmk) relation. 
Panels e) and f): Same as panels a) and b), but for the PW($K$,\rmk) relation. 
Panels g) and h): Same as panels a) and b), but for the PW($K$,\imk) relation. 
Panels i) and j): Same as panels a) and b), but for the PW($K$,\jmk) relation. 
}\label{fig4}
\end{figure*}

Figures \ref{fig3} and \ref{fig4} show the 
observed optical-NIR and NIR PW relations. The coefficients of the PW 
relations and their dispersions are listed in Table~\ref{tab:table_plw_nir}. 
The trend is similar to the optical PW relations: the dispersion decreases 
with increasing difference in the central wavelengths of the adopted color.
% EP 
The above empirical evidence is the consequence of two independent mechanisms. 
i) The increased difference in central wavelength causes a substantial 
change in the color coefficient, and indeed it ranges from 4.92 PW($R$,\vmr) to 
0.11 PW($K$,\bmk). 
In particular, the optical--NIR PW relations have color coefficients that 
are systematically smaller than unity. This means that they are less affected 
by uncertainties in the mean colors. Note that the PW($I$,\bmi) relation 
is also characterized by a color coefficient that is smaller than unity, but 
the errors in the coefficients of this relation 
are on average larger when compared with optical--NIR PW relations.   
ii) The increased difference in central wavelength implies a stronger 
sensitivity to the mean effective temperature, and in turn a more uniform 
distribution of RRLs across the instability strip and along the 
PW relation. 

The dispersion of FU pulsators in the NIR PW relation (panels i) and j) of 
Fig. \ref{fig4})  
is larger than for the FO PW relations. Although the latter are fainter
they also have smaller luminosity amplitudes, and their mean NIR colors are on 
average more accurate. This is the reason why we did not include the PW($H$,\jmh) 
and the PW($K$,\hmk) relations, the \jmh\ and the \hmk\ colors being less accurate. 

\begin{figure*}[tbp]
\centering
\includegraphics[width=13cm,height=19cm]{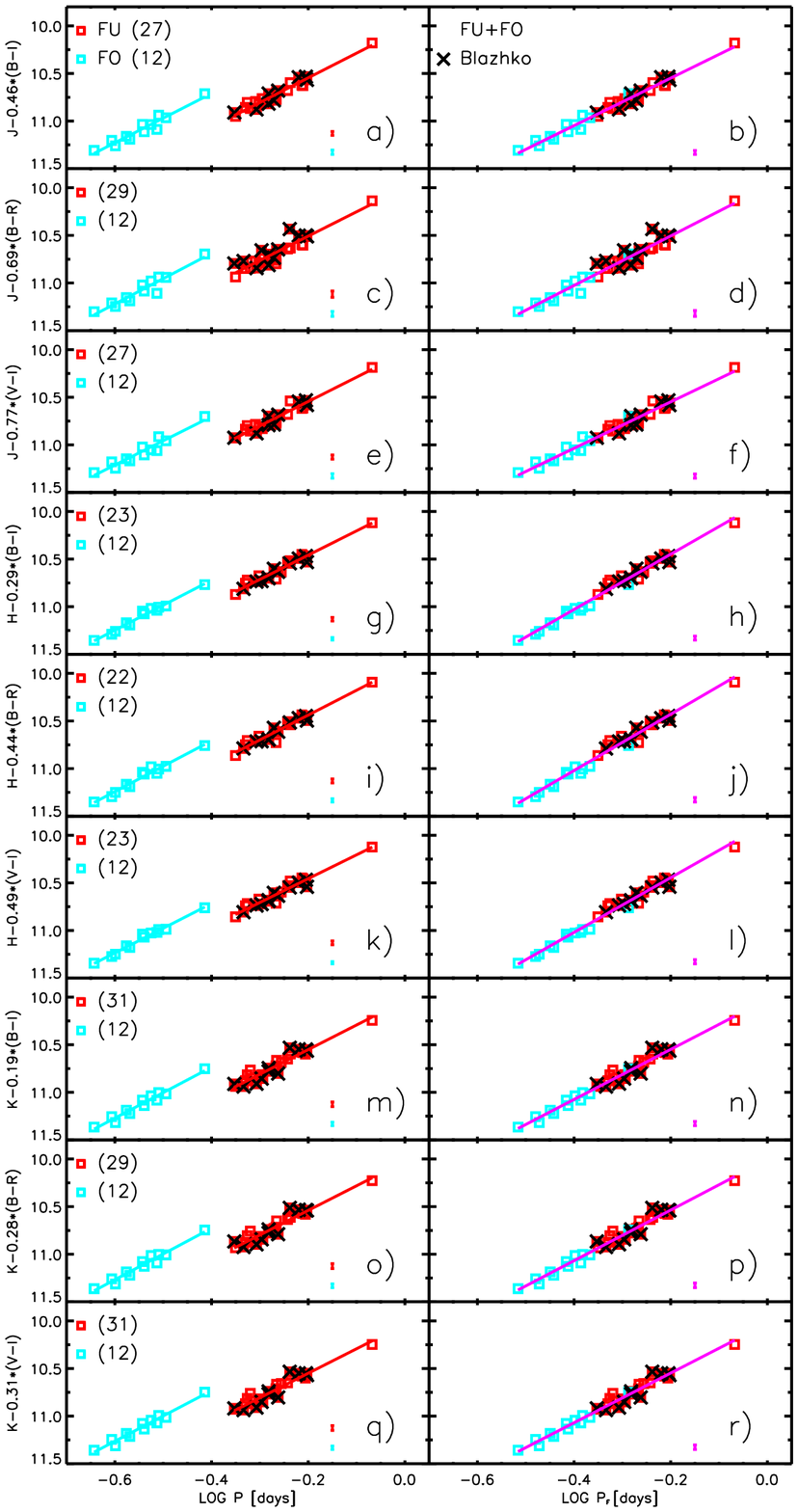}
\vspace*{0.5truecm} 
\caption{From top to bottom observed optical--NIR three-band PW relations 
for RR Lyrae in M4. Symbols and lines are the same as in Fig.~1. 
a): PW($J$,\bmi) relation for FU and FO pulsators. 
b): Same as panel a), but for the entire sample of RR Lyrae. The 
periods of FO variables were fundamentalized using the relation: 
$\log  P_{FU}$= $\log  P_{FO}$+0.127.    
c) and d): Same as panels a) and b), but for the PW($J$,\bmr) relation. 
e) and f): Same as panels a) and b), but for the PW($J$,\vmi) relation. 
g) and h): Same as panels a) and b), but for the PW($H$,\bmi) relation. 
i) and j): Same as panels a) and b), but for the PW($H$,\bmr) relation. 
k) and l): Same as panels a) and b), but for the PW($H$,\vmi) relation. 
m) and n): Same as panels a) and b), but for the PW($K$,\bmi) relation. 
o) and p): Same as panels a) and b), but for the PW($K$,\bmr) relation. 
q) and r): Same as panels a) and b), but for the PW($K$,\vmi) relation. 
}\label{fig5}
\end{figure*}

To fully exploit the power of the current optical and NIR RRLs mean magnitudes
for the distance of M4 we also adopted the three-band PW relations.
These are PW relations in which the pass-bands adopted in the color index differ
from the pass-band adopted for the magnitude.
Among all the possible combinations we only selected PW relations 
in which the coefficient of the color index is smaller than unity. 
Fig.~\ref{fig5} shows the nine empirical PW relations we 
selected, while Table~\ref{tab:table_plw_nir} gives the coefficients, their uncertainties, and the 
standard deviations. The data listed in Table~\ref{tab:table_plw_nir} indicate that both the 
uncertainties in the coefficients and the standard deviations of the 
optical-NIR, three-band PW relations \citep{riess2011} 
are up to a factor of two smaller than in the optical-NIR, two-band 
ones.

%_______________________________________________________________________________
\section{Theoretical Period--Luminosity--Metallicity and 
Period--Wesenheit--Metallicity relations}\label{chapt_predictions}

To estimate the distance to M4 we adopted theoretical PLZ and PWZ relations. 
The reasons are twofold. i) The five field RRLs for which
accurate trigonometric parallaxes are available do not yet have accurate optical and 
NIR mean magnitudes \citep{bene2011}. The only exception is 
RR Lyr itself for which accurate {\it BV\/} \citep{szeidl1997} and NIR 
\citet{sollima2008} mean magnitudes are available in the literature.     
ii) Absolute distances based on predicted PL relations for 
RRLs do agree quite well with similar cluster distances 
based on solid distance indicators 
\citep{cassisi2004,delprincipe2006,sollima2006,bono2011,coppola2011}.

To provide a detailed theoretical framework for both PLZ and PWZ relations 
we adopted the large set of nonlinear, convective pulsation models recently 
computed by Marconi et al. (2014, in preparation). Models were computed for 
both fundamental and first overtone pulsators and cover a broad range in  metal abundance  
(--2.62$\le$[Fe/H]$\le$--0.29). The stellar masses and the luminosity 
were fixed by using evolutionary prescriptions for HB 
models provided \citep{pietrinferni2004,pietrinferni2006}. The reader 
interested in more details concerning the physical and numerical assumptions   
adopted to construct the pulsation models is referred to 
\citep[][ Marconi et al. 2014, in preparation and references therein]{bono1994}.  
To take account of the metallicity dependence Marconi et al. (2014, in preparation) performed 
a linear fit of FU and FO pulsators including a metallicity term. 
In this context it is worth mentioning that the coefficients of the 
metallicity term in the PLZ relations attain very similar values in 
both the optical and NIR bands. 

% Referee 7_ii
The theoretical 
predictions were transformed into the observational plane by adopting bolometric 
corrections and color-temperature relations provided by \citep{castelli1997a,castelli1997b}. In 
passing we note that above transformations are based on static, LTE atmosphere models. The use 
of static atmosphere models in dealing with the atmospheres of variable stars has already 
been addressed by \citep{bonocaputostell1994}. The non-LTE effects have impact on individual lines 
of individual elements, but they minimally affect broad-band colors of RRLs \citep[see 
Fig.~11 in ][]{kudritzki1979}.

% Referee 7_i 
The HB evolutionary models we adopted to compute the mass-luminosity relation do not 
take account of any possible rotation. The empirical scenario concerning the rotational velocity of 
RRLs is 
far from being settled. Dating back almost twenty years ago, 
in a seminal investigation using roughly two dozen field RRLs \citet{petersoncarneylath1996} found an upper limit to 
the equatorial rotational velocity of $V_{rot}sin$i $<$ 10 $\kms$. A more tight constraint 
was recently provided by \citet{prestonchadid2013} using thousands of high-resolution 
spectra for three dozen field RRLs. They found an upper limit $V_{rot}sin$i $<$ 6 $\kms$. 
The above findings suggest that rotation plays a minor, if any, role in shaping the atmospheric 
properties of RRLs.

\begin{figure*}[tbp]
\centering
\includegraphics[width=14cm,height=15cm]{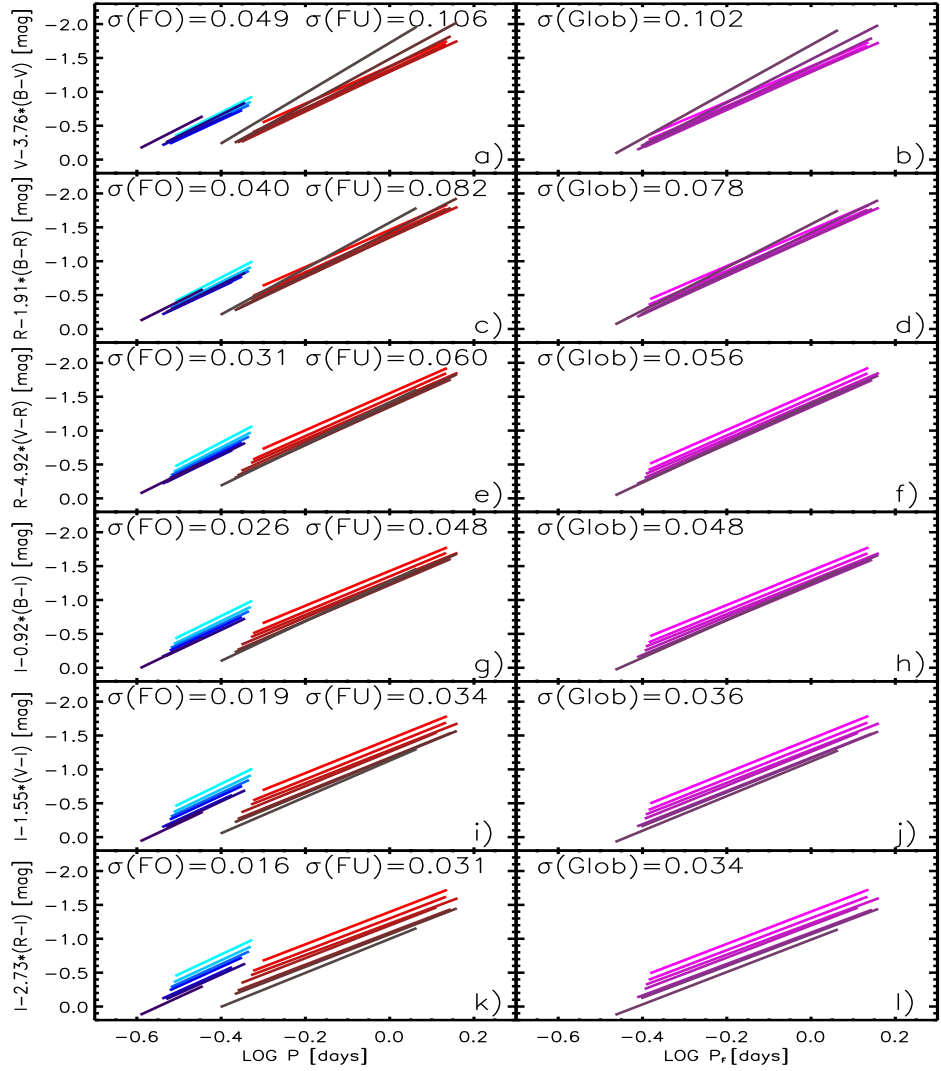}
\vspace*{0.5truecm} 
\caption{From top to bottom predicted optical PW relations for RR Lyrae models. 
The PW relations were estimated by using the reddening law for M4 provided by \citep{hendricks2012}
Panel a): Lines of different colors display PW relations for FU and FO pulsators. 
The PW($V$,\bmv) relations range in metallicity from [Fe/H]=-2.62  
(brighter) to [Fe/H]=-0.29 (fainter). See Table~\ref{tab:table_dis} for more details concerning 
the adopted metallicities.  
Panel b): Same as panel a), but for the entire sample of RR Lyrae models. 
The periods of FO models were fundamentalized using the relation: 
$\log  P_{FU}$= $\log  P_{FO}$+0.127.
Panels c) and d): Same as panels a) and b), but for the predicted PW($R$,\bmr) relation. 
Panels e) and f): Same as panels a) and b), but for the predicted PW($R$,\vmr) relation. 
Panels g) and h): Same as panels a) and b), but for the predicted PW($I$,\bmi) relation. 
Panels i) and j): Same as panels a) and b), but for the predicted PW($I$,\vmi) relation. 
Panels h) and k): Same as panels a) and b), but for the predicted PW($I$,\rmi) relation. 
}\label{fig6}
\end{figure*}

\begin{figure*}[tbp]
\centering
\includegraphics[width=13cm,height=15cm]{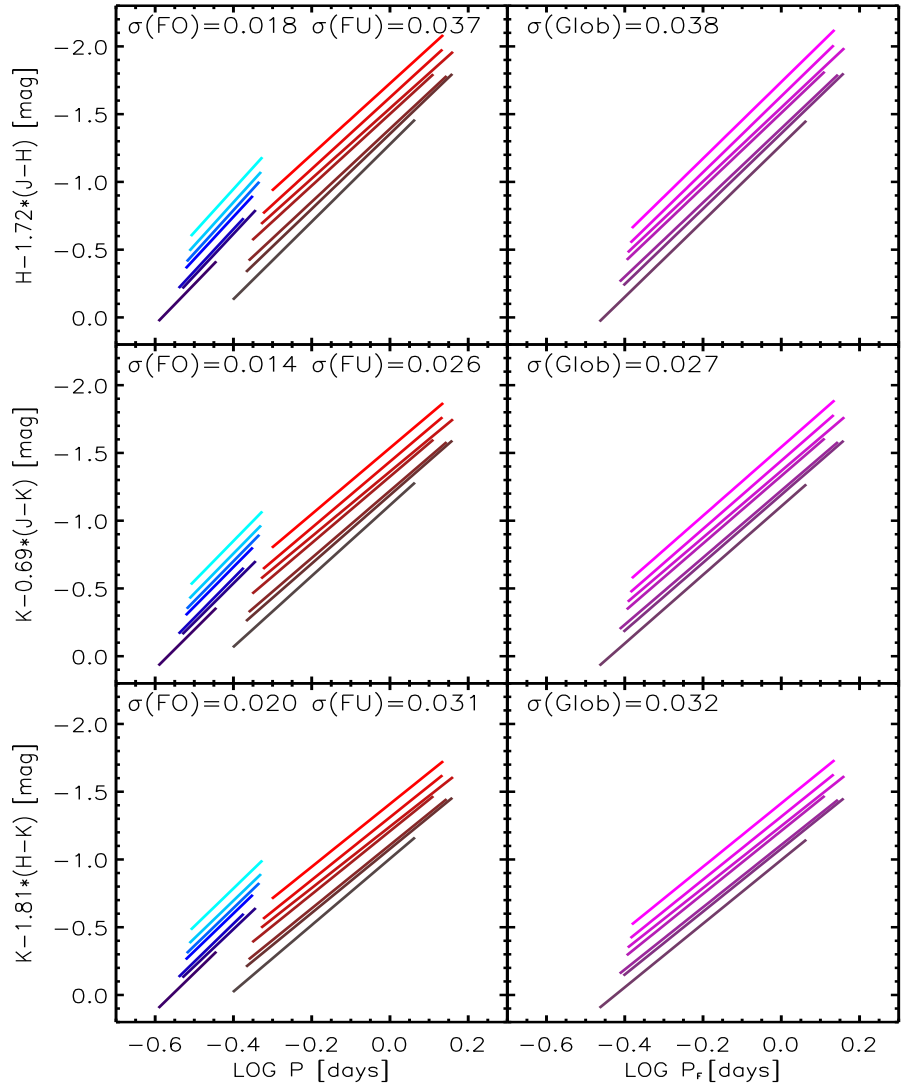}
\vspace*{0.5truecm} 
\caption{Same as Fig.\ref{fig6}, but for the predicted 
NIR PW relations. 
}\label{fig7}
\end{figure*}

The PWZ relations were  computed following the same approach adopted for the 
PLZ relations. Figures~\ref{fig6} and \ref{fig7} display 
optical and NIR PW relations. The coefficients, their errors 
and the standard deviations are listed in Table~\ref{tab:table_plw_th_nir}.

The theoretical framework concerning the PWZ relations shows several interesting 
features. 
The optical Wesenheit magnitudes---$W$($V$,\bmv), $W$($R$,\bmr)---display a 
peculiar trend with metallicity. An increase in metallicity from Z=0.0001 to 
Z=0.001 makes {\it W\/}, at fixed period, fainter while for still more metal-rich structures 
it becomes brighter. This is the reason why the coefficients of the metallicity 
term attain, for the quoted PWZ relations, vanishing values and why their 
standard deviations are larger. The ranking with the metallicity becomes 
linear for the PWZ($R$,\bmr) relation and increases for the PWZ relations including 
the $I$ band. The coefficient of the metallicity term attains, once again, very 
similar values in both the optical-NIR and NIR PWZ relations.     

\begin{figure*}[tbp]
\centering
\includegraphics[width=13cm,height=19cm]{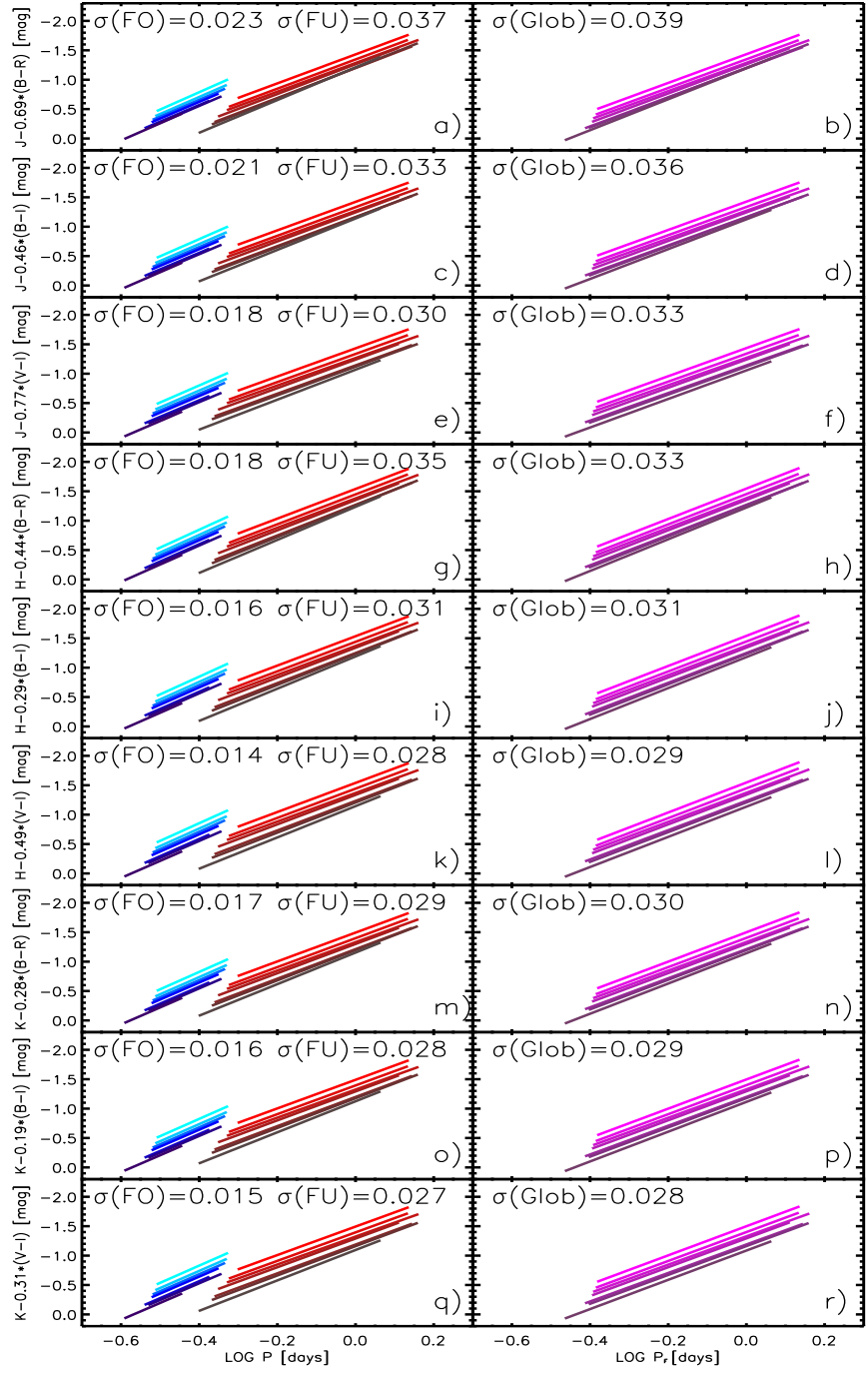}
\vspace*{0.5truecm} 
\caption{Same as Fig.\ref{fig6}, but for the predicted 
optical-NIR three-band PW relations. 
}\label{fig8}
\end{figure*}

Fig.~\ref{fig8} shows the nine predicted optical-NIR, three-band PW relations, 
while Table~\ref{tab:table_plw_th_nir} gives the coefficients, their uncertainties and the standard 
deviations. The current predictions indicate that both the uncertainties in the 
coefficients and the standard deviations of the three-band PWZ relations 
attain similar values when compared with two-band PWZ relations. This 
evidence suggests that the improvement in the empirical three-band 
PWZ relations might be a consequence of the fact that the optical mean 
colors adopted in the three-band PWZ relations are more precise 
than the optical-NIR mean colors adopted in the two-band ones.   

\begin{figure*}[tbp]
\centering
\includegraphics[width=13cm,height=13cm]{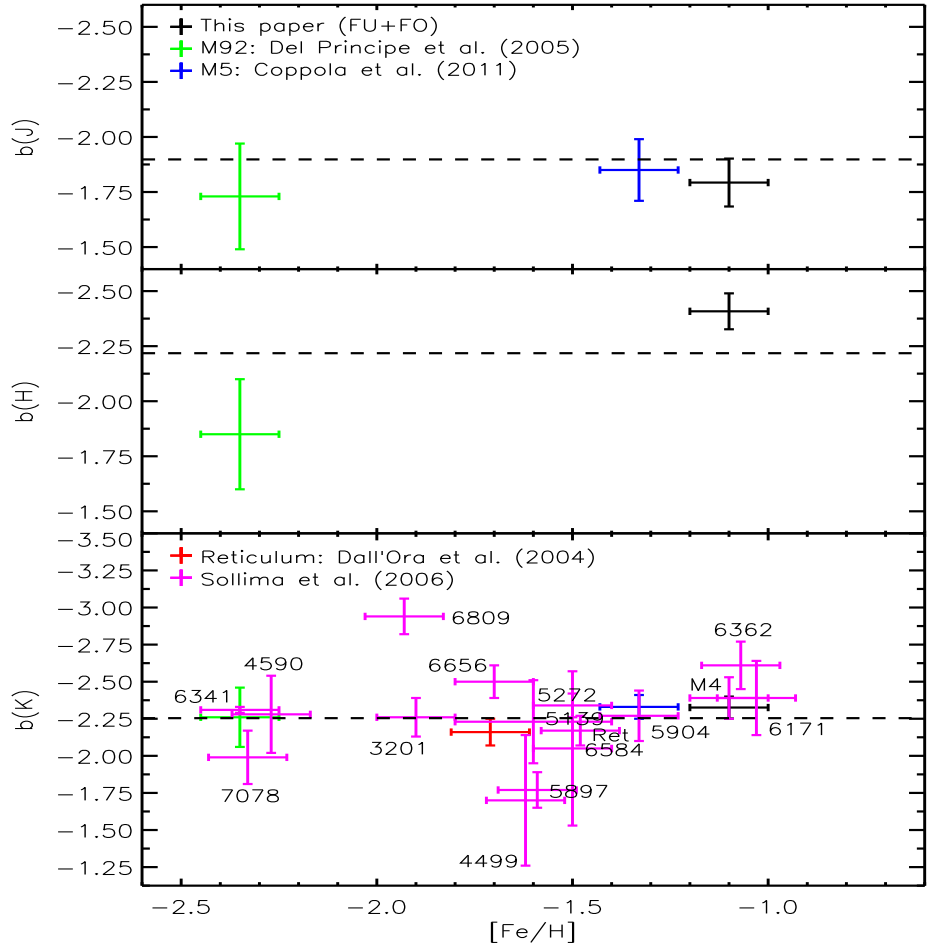}
\vspace*{0.5truecm}
\caption{From top to bottom slopes of observed NIR PL relations
as a function of the iron abundance. The iron abundances are based
on the GC metallicity scale provided by \citet{carr09}.
Top: slope of the $J$--band PL relation. The error bars display
the error on the slope of the PL relations and the uncertainty on
the metal abundance. The black line shows the slope of the predicted 
PLZ relation.  
Middle: Same as the top, but for the $H$--band PL relation.
Bottom: Same as the top, but for the $K$--band PL relation.
\label{fig9}
}
\end{figure*}

To further constrain the theoretical framework adopted to estimate the distance 
to M4, we performed a detailed comparison between the slope of NIR PL relations 
available in the literature and the slope of the current PLZ relation. Data 
plotted in the bottom panel of Fig.~\ref{fig9} indicate good agreement 
between the predicted $K$-band slope (dashed line), the observed slope for M4 
and similar estimates for Galactic globulars. The agreement appears quite good, 
within the errors, over the entire metallicity range. The same applies for the 
$J$ band (top panel), but theory and observations agree within $\sim$1 $\sigma$.
We cannot reach a firm conclusion concerning the $H$ band, since only two 
% Referee 5_i 
empirical estimates are available and they attain intermediate values.

%_______________________________________________________________________________
\section{Distance determinations to M4 based on empirical and predicted 
Period--Luminosity and Period--Wesenheit relations}\label{chapt_distances}

% Referee 9_ii 
%_____________________________________________________________________________________
\subsection{Cluster distances based on the absolute distance of RR Lyr itself}

Thanks to the use of the Fine Guidance Sensor on board the \hst, \citet{bene2011} provided 
accurate estimates of the trigonometric parallaxes for five field RRL: SU Dra, XZ Cyg, RZ Cep, 
XZ Cyg and RR Lyr. 
To fix the zero-points of the empirical relations we decided to use RR Lyr itself. The reasons 
are the following: 

i) The absolute parallax to RR Lyr itself is the most precise (3.77$\pm$0.13 \mas) among the 
calibrating RRLs and an accurate estimate of the reddening toward RR Lyr is also available 
\citep[$E(\bmv)$=0.02$\pm$0.03; ][]{sollima2008}.
ii) Accurate optical \citep[{\it BV\/}; ][]{szeidl1997} and NIR \citep[{\it JHK\/}; ][]{sollima2008}
mean magnitudes are available in the literature (see Table~\ref{tab:table_rrlyr}).  
Note that RR Lyr together with UV Oct and XZ Cyg are affected by the Blazhko effect. 
This means that the typical uncertainty on its mean {\it BV\/} magnitudes is of the order 
of 0.10 mag.   
iii) An accurate estimate of the iron abundance is also available
\citep[\lbrack Fe/H\rbrack =-1.41$\pm$0.13; ][]{kolenberg2010}.
To provide a homogeneous metallicity scale with RRLs in M4 
we took account of the difference in the adopted solar iron 
abundance between \citet{carr09} and \citet{kolenberg2010} 
([Fe/H]$\sim$-1.50$\pm$0.13, see Table~\ref{tab:table_rrlyr}).
The above iron abundance, once transformed from the \citet{zinnwest1984} to
the \citet{carr09} metallicity scale, indicates that RR Lyr is $\sim$0.40 dex 
more metal--poor than RRLs in M4 (see Table~\ref{tab:table_rrlyr}).

To provide an empirical estimate of the absolute distance to M4, we used the slopes of 
both PL and PW relations listed in Tables~2,3 and 4. The extinction corrections to the 
observed mean magnitudes have been estimated using the \citet{cardelli} semi-empirical 
reddening law. The apparent magnitudes have also been corrected for the difference in 
iron abundance ($\Delta$ [Fe/H]=--0.40) between M4 and RR Lyr using the metallicity 
coefficients of the predicted PLZ and PWZ relations (see \S~7.2). The absolute distances 
based on the NIR PL relations listed in Table~\ref{tab:table_dmod_obs_pl} 
give a weighted mean distance modulus 
of 11.35$\pm$0.03$\pm$0.05 mag (FU) and 11.35$\pm$0.03$\pm$0.05 mag (FO$+$FU). 
The former error is the error on the mean, while the latter is the standard deviation.  
The two estimates are, within the errors, identical. They also agree, within 1$\sigma$, 
with accurate estimates available in literature (see Table~1) and with distances based 
on predicted PLZ and PWZ relations (see \S~7.2).   
The RR Lyr zero-point was also used for the eight optical-NIR PW relations 
including {\it BVJHK\/} bands. The weighted mean distance modulus is 
11.31$\pm$0.02$\pm$0.07 mag (FU) and 11.32$\pm$0.02$\pm$0.07 mag (FO$+$FU), 
respectively. The weighted mean distance modulus only based on optical-NIR 
and on NIR PW relations is 11.31$\pm$0.02$\pm$0.06 mag (FU) and 
11.32$\pm$0.02$\pm$0.06 mag (FO$+$FU). 
The above estimates agree quite well with similar estimates available in the 
literature and within 1$\sigma$ with distances based on predicted PWZ relations.

%_____________________________________________________________________________________
\subsection{Cluster distances based on predicted PLZ and PWZ relations}

We estimated the true distance modulus of M4 adopting the 
predicted PLZ relations discussed in the \S~\ref{chapt_predictions}
together with the mean reddening ($E(\bmv)=0.37\pm0.01$ mag), 
the reddening law provided by H12 and a mean metal 
abundance of [Fe/H]=--1.10. The latter is a mean value based 
on iron abundances provided by \citet[][\lbrack Fe/H\rbrack=--1.13]{marino2008},
\citet[][\lbrack Fe/H\rbrack=--1.18]{carr09}
and by \citep{malavolta2014} using both RGB stars ([Fe/H]=--1.07) 
and main sequence stars and ([Fe/H]=--1.16). Note that to provide homogeneous iron 
abundances the above measurements were rescaled to the same solar 
iron abundance adopted by \citet{carr09}. Moreover,
to estimate individual distances we are using the mean apparent 
magnitudes together with the zero-point and the slope of the predicted 
PLZ relations.

\begin{figure*}[tbp]
\centering
\includegraphics[width=12cm,height=14cm]{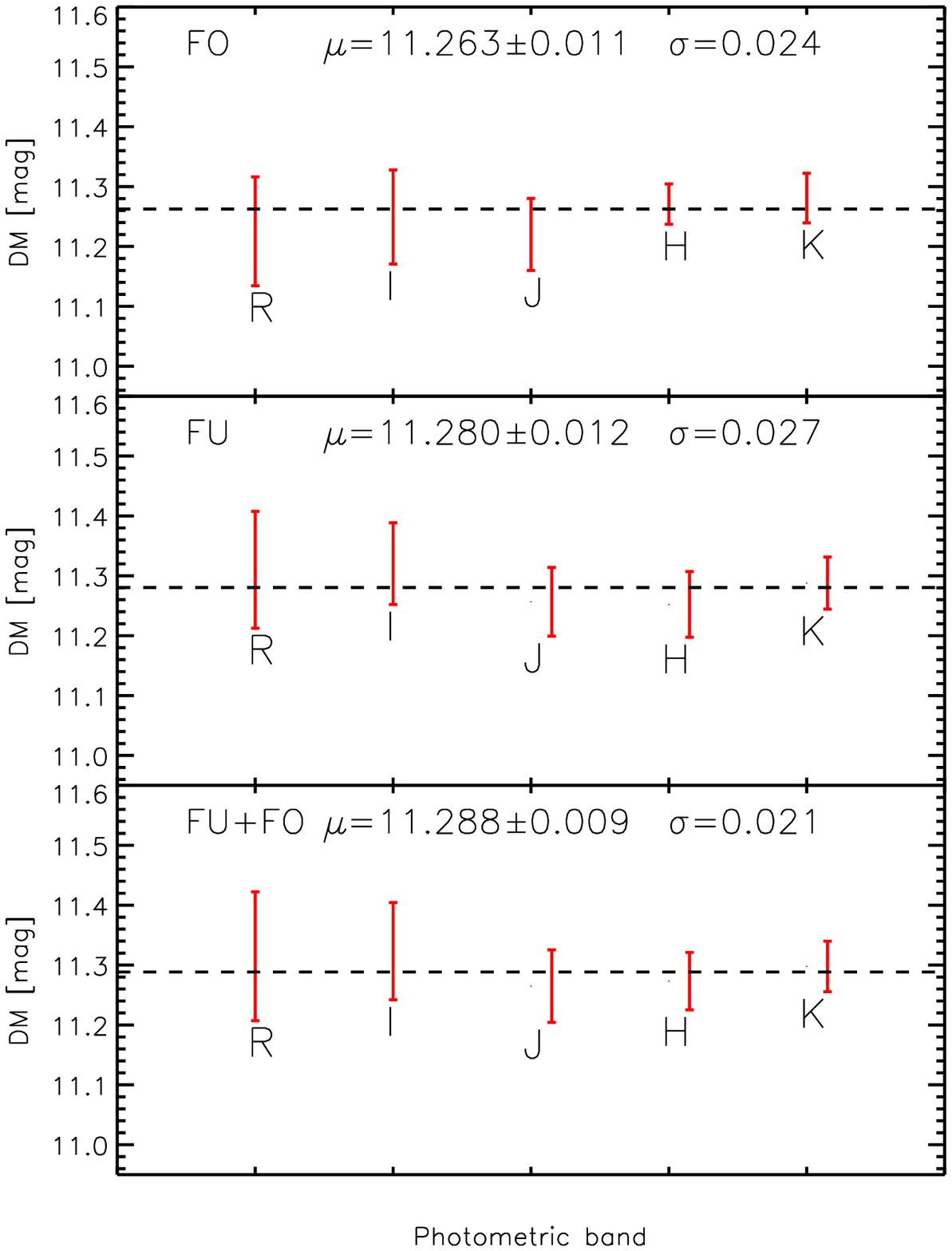}
\vspace*{0.5truecm} 
\caption{Top: True DM based on optical and NIR predicted 
FO PL relations that take account of the metallicity dependence. Following \citet{carr09} 
we adopted for M4 an iron abundance of [Fe/H]=-1.10 The error bars include 
the photometric error, the extinction error, and the standard deviation 
of the adopted PL relation. The dashed lines shows the weighted mean true 
DM ($\mu$). The error on the mean and the standard deviations 
are also labeled.  
Middle: Same as the top, but for optical and NIR predicted FU PL relations.
Bottom: Same as the top, but for optical and NIR predicted PL relations including 
FU and FO RR Lyrae models.    
}\label{fig10}
\end{figure*}

The results for true distance moduli are listed in Table~\ref{tab:table_dmod_th_pl} and plotted in 
Fig.~\ref{fig10}.
% Referee 8-bis.ii 
The error on the distance modulus takes account of the photometric error, for uncertainties 
in the mean reddening ($E(\bmv)=0.37\pm0.10$ mag)), in the total-to-selective extinction ratio 
($R_V$=3.62$\pm$0.07), in the mean metallicity ($\sigma$([Fe/H])=0.1 dex) and for the standard 
deviation of the adopted PLZ relation.
The weighted true distance moduli based on FU, FO and on the entire sample 
of variables agree within 1$\sigma$. However, the distance modulus based 
on FOs attains a smaller value compared with the FUs and with the combined 
sample (see labeled values). The main culprits are distance determinations 
based on optical PLZ relations, and indeed if we only use the NIR PLZ relations 
we find $\mu=$ 
11.266$\pm$0.014 (error on the mean) $\pm$0.025 (weighted standard deviation) mag 
for FOs, 
11.271$\pm$0.012 $\pm$0.020 mag for FUs and 
11.283$\pm$0.010 $\pm$0.018 mag for the entire sample.    

Note that the weighted standard deviation of the true distance moduli 
based on the $H$-band FO PLZ relations is smaller than in the $J$ and 
$K$ bands. The difference is due to the fact that the observed standard 
deviation in the $H$ band is smaller when compared with the $J$ and $K$ bands. 

The true distance moduli based on optical, optical-NIR and NIR PWZ relations 
show a more complex trend. The optical PWZ relations with vanishing metallicity 
terms display a large scatter when compared with true distance moduli based on the 
other NIR PWZ relations. In passing we note that the scatter of the true distance 
moduli based on the optical PWZ relations slightly decreases when using PW relations 
that neglect the metallicity dependence.   

\begin{figure*}[tbp]
\centering
\includegraphics[width=12cm,height=14cm]{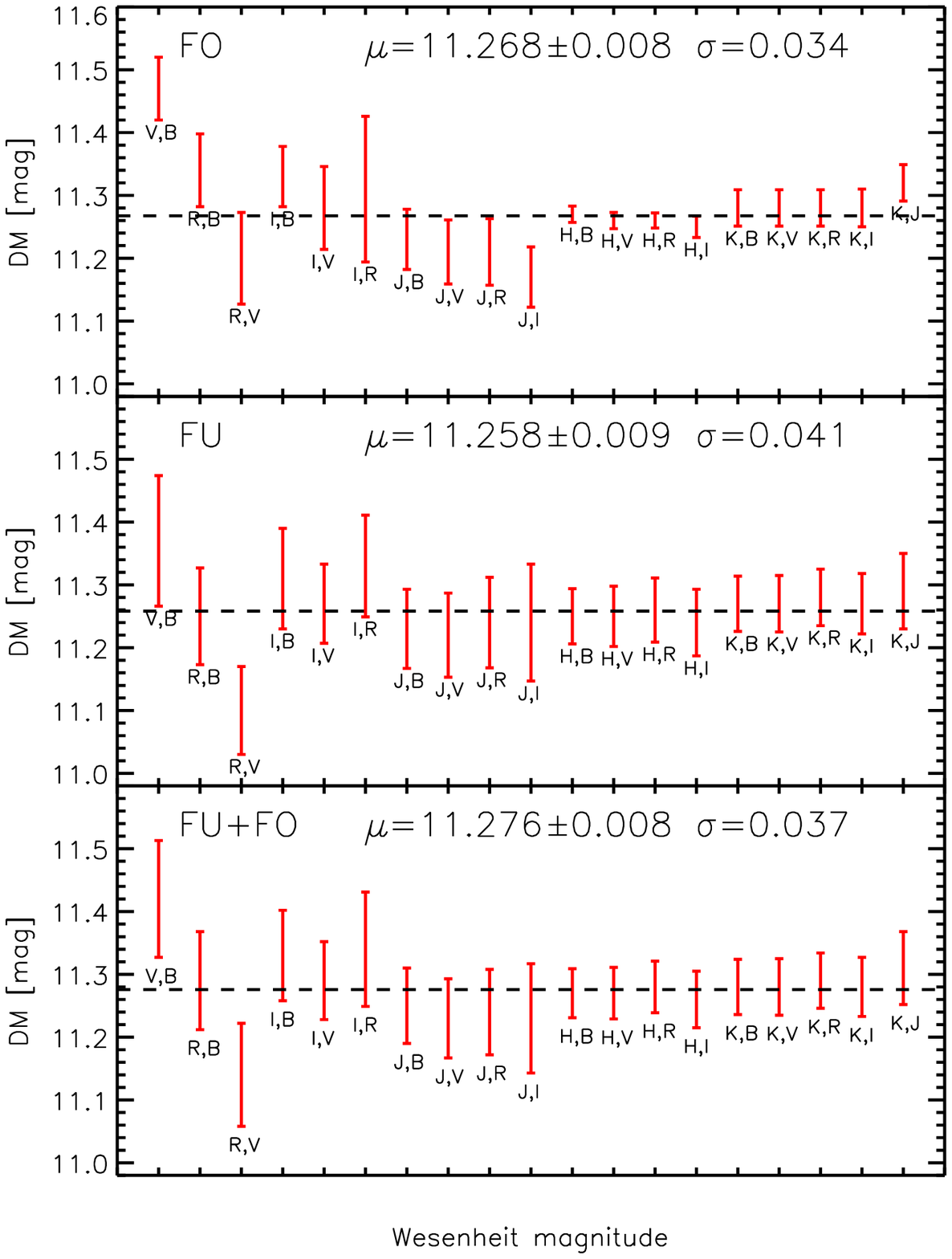}
\vspace*{0.5truecm} 
\caption{Top: True DM based on optical and NIR predicted 
FO PW relations that take account of the metallicity dependence. Following \citet{carr09} 
we adopted for M4 an iron abundance of [Fe/H]=-1.10 The error bars include 
the photometric error, the extinction error, and the standard deviation 
of the adopted PW relation. The dashed lines shows the weighted mean true 
DM ($\mu$). The error on the mean and the standard deviations 
are also labeled.  
Middle: Same as the top, but for optical and NIR predicted FU PW relations.
Bottom: Same as the top, but for optical and NIR predicted PW relations including 
FU and FO RR Lyrae models.    
}\label{fig11}
\end{figure*}

Among the true distance moduli based on optical-NIR PWZ relations those 
including the $J$ band attain slightly smaller values 
(see Table ~\ref{tab:table_dmod_th_plw_nir} and Fig.~\ref{fig11}). 
The reason for the difference it is not clear. The adopted color-temperature relations 
to transform the theoretical models into the observational plane might be 
a possible culprit. A similar difference was also found in optical and 
in optical-NIR color-magnitude diagrams by \citet{bono2010}.
The weighted standard deviations of the true distance moduli of FO 
PWZ relations including the $H$ band are smaller than in the 
$J$ and $K$ bands. The reasons for the difference are the same 
as for the PLZ relation.    
Interestingly enough, the true distance moduli based on optical-NIR 
PWZ relations including the $H$ and $K$ bands are very accurate 
and display a very small dispersion. 

The weighted mean true distance modulus based on optical, optical-NIR and NIR PWZ relations 
agrees within 1$\sigma$. The agreement between the three different sets of distance 
determinations minimally improves if we only use optical-NIR and NIR PWZ relations: 
11.263$\pm$0.006$\pm$0.021 mag for FOs, 
11.259$\pm$0.005$\pm$0.019 mag for FUs and 
11.272$\pm$0.005$\pm$0.019 mag for the entire sample.   
The lack of a clear dependence of the estimated distance moduli on the photometric 
bands is further supporting the accuracy of the reddening law adopted to estimate 
the PW relations and the true mean optical and NIR magnitudes \citep{pietrzynski2006}. 

\begin{figure*}[tbp]
\centering
\includegraphics[width=12cm,height=14cm]{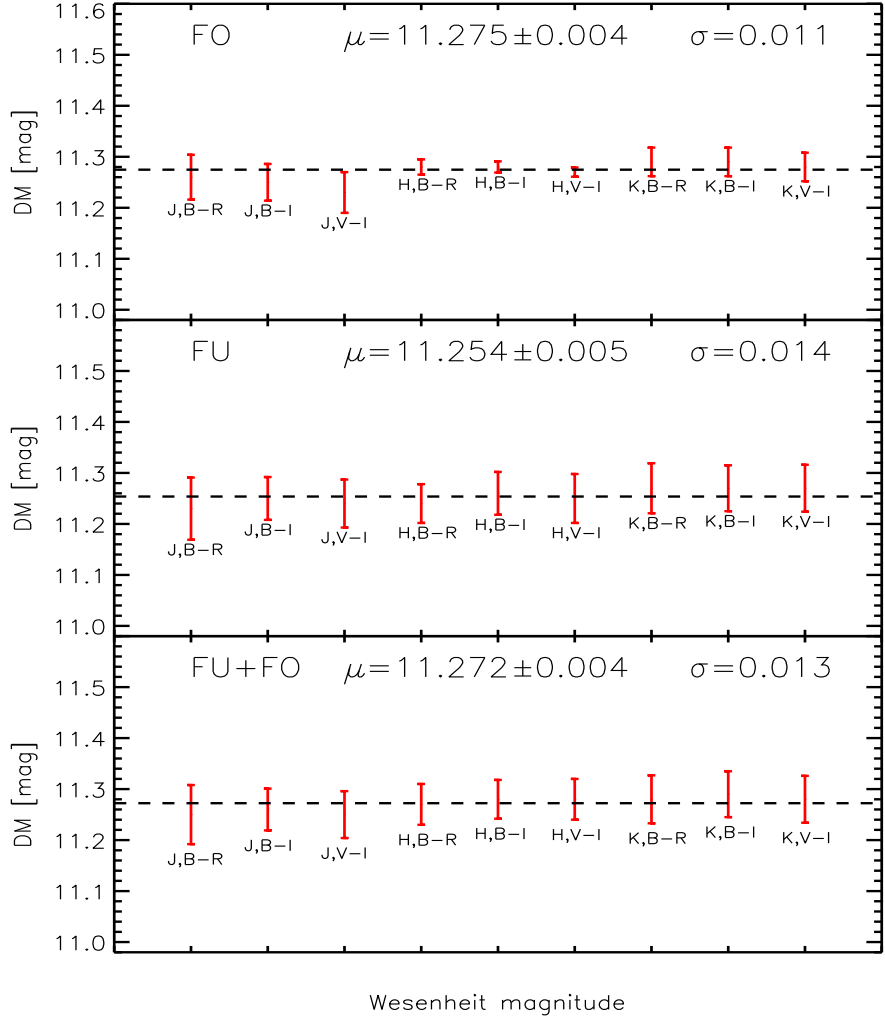}
\vspace*{0.5truecm} 
\caption{Same as Fig.~\ref{fig11}, but the true distance moduli were estimated by adopting 
optical--NIR, three-band PWZ relations. The optical color adopted in the PWZ relation is independent of the adopted NIR magnitude.  
}\label{fig12}
\end{figure*}

Distance moduli based on three-band PWZ relations have, as expected, a smaller
dispersion when compared with two-band ones (see Figure~\ref{fig12}). 
In particular, the weighted mean true distance modulus based on FOs is  
11.275$\pm$0.004$\pm$0.011 mag for FOs, while those based on FUs is 
11.254$\pm$0.005$\pm$0.014 mag for FUs and the those one based on the entire 
sample is 11.272$\pm$0.004$\pm$0.013 mag.

The current distance evaluations based on predicted PLZ and PWZ relations agree 
with each other within 1$\sigma$. They also agree quite well with distance 
determinations to M4 based on solid standard candles, and in particular, with 
the distance recently provided by \citet{kaluzny2013}, who obtained a value 
of 11.30$\pm$0.05 mag using three eclipsing binaries. The same outcome applies to the recent 
distance evaluation based on the HB luminosity level (11.28$\pm$0.06 mag) 
provided by H12. The above findings indicate that NIR PLZ relations and optical-NIR/NIR 
PWZ relations can provide individual distances to GCs hosting a good sample of RRLs
with a precision better than 2-3\%.

%_______________________________________________________________________________
\section{Final remarks and conclusions}\label{chapt_conclusion}

We have presented new and precise optical, optical-NIR and NIR PL 
relations and PW relations. We have provided 
independent empirical relations for first overtone, fundamental and for the entire 
sample of RRLs in M4.   

The mean weighted visual apparent magnitude of 44 cluster RRLs is 
$\left<V\right>=13.329\pm0.000\pm0.177$ mag, where the former error refers 
to the error on the mean and the latter to the weighted standard deviation. 
The current estimate agrees quite well with similar evaluations available 
in the literature. Indeed, \citet{liujanes1990b} using four fundamental
variables (V2, V15, V32, V33) found $\left<V\right>=13.287\pm0.025\pm0.213$ mag, 
while \citet{clementini1994} using four fundamental variables 
(V2, V15, V29, V42) found $\left<V\right>=13.371\pm0.001\pm0.139$ mag.  
This is a relevant stepping stone for the forthcoming investigation in which we 
plan to estimate the absolute \citep{bono2010} and the relative 
\citep{vandenberg2013} age of M4 using both optical and NIR photometry.  

%Referee 9 
We have estimated the true distance modulus to M4 using the observed slopes 
and RR Lyr itself to fix the zero-point. RR Lyr is, out of the five field RRLs 
with accurate trigonometric parallaxes measured by \hst~\citep{bene2011},  
the calibrator with the most precise distance and with both optical ({\it BV\/}) and 
NIR ({\it JHK\/}) mean magnitudes.     
Moreover, accurate estimates of both the iron content and the reddening are 
also available. The main drawback in using RR Lyr is that it is affected by 
the Blazhko effect together with UV Oct and XZ Cyg. The impact is minimal 
in the NIR bands, but the uncertainty in the mean optical bands is of the 
order of 0.10 mag. 
To determine the true distance modulus we took account of the difference 
in iron abundance between RR Lyr and M4 ($\Delta$[Fe/H]$\sim$0.40, according 
to the metallicity scale by \citet{carr09}). The difference was estimated 
using predicted PLZ and PWZ relations.   
The weighted mean true distance modulus based on three independent empirical 
NIR PL relations and the entire sample of RRLs (FO$+$FU) is 11.35$\pm$0.03$\pm$0.05 mag.        
The weighted mean true distance modulus based on eight different empirical optical--NIR 
and NIR PW relations is 11.32$\pm$0.02$\pm$0.07 mag (FO$+$FU).
The above estimates agree quite well with similar estimates avaialble in the
literature and within 1$\sigma$ with distances based on predicted PLZ and 
PWZ relations.

We also estimated the true distance moduli to M4 using predicted optical, optical-NIR 
and NIR PLZ and PWZ 
relations. The theoretical relations are based on a broad range of nonlinear, convective 
pulsation models for RRLs. They were constructed for both FO and FU pulsators 
and cover a broad range in stellar masses (M=0.80-0.55$\msun$) and  metal 
abundances (Z=0.0001--0.02).    

The true distance moduli based on the PLZ relations take account of uncertainties in the 
mean reddening, in the photometry, in the mean metallicity and in the standard deviation 
of the adopted PLZ relation. The true distance moduli based on the PWZ relations 
take account of uncertainties in the photometry and in the mean metallicity.

We found that true distance moduli based on NIR PLZ relations are, as expected, 
characterized by smaller intrinsic dispersions when compared with optical 
PLZ relations. The difference is mainly the consequence of steeper slopes in 
the PLZ relation at longer wavelengths.     

We also found that optical PWZ relations present larger intrinsic dispersions 
when compared with optical-NIR and NIR PWZ relations. The difference is mainly 
the consequence of a nonlinear dependence on the metallicity in the optical 
regime when compared with the optical-NIR and with the NIR regimes.   

True distance moduli based on FO PLZ and PWZ relations display smaller weighted 
standard deviations when compared with FU, and FO$+$FU PLZ and PWZ relations. 
This evidence---taken at face value---seems to argue in favor of the idea 
that FO variables can provide accurate and precise individual distance moduli. 
However M4 hosts a dozen of FO RRLs, further support based on GCs hosting 
sizable samples of FO pulsators is required.

Recent findings based on MIR photometry collected with WISE 
of field RRLs strongly suggest a very small intrinsic scatter in 
the PL relation and a mild dependence on the metal abundance \citep{madore2013,klein2014}. 
The use of MIR photometry collected 
with Spitzer within the Carnegie RR Lyrae Project (CRRP) appears as a natural   
development of this investigation. The key advantage in this approach is that 
we can use the five empirical calibrators for which are available accurate 
MIR mean magnitudes.  

The above findings appear very promising not only for the next generation 
of extremely large telescopes (ELTs), namely the European-ELT 
[E-ELT]\footnote{http://www.eso.org/public/teles-instr/e-elt.html},
the Thirty Meter Telescope [TMT]\footnote{http://www.tmt.org/}, 
and the Giant Magellan Telescope [GMT]\footnote{http://www.gmto.org/}, 
but also for James Web Space Telescope [JWST]\footnote{http://www.jwst.nasa.gov/} 
and EUCLID\footnote{http://sci.esa.int/euclid/}.
Future ground-based and space facilities will be equipped with a suite of 
NIR and MIR detectors to perform accurate photometry and spectroscopy of 
old stellar tracers in the nearby Universe. This is an unique opportunity to improve 
the cosmic distance scale of stellar systems hosting old stellar populations 
(early and late type galaxies) from the Local Group to the Virgo galaxy cluster.  

%______________________________________________________________________________
\acknowledgments
It is a pleasure to thank the organizers of the MIAPP--Munich Institute 
for Astro and Particle Physics--workshop on {\ em The Extragalactic 
Distance Scale}. During the workshop several authors have had the 
opportunity to present the preliminary results of this investigation 
and to discuss pros and cons of the RR Lyrae distance scale in a very 
fruitful and pleasant environment.
In addition we warmly thank R.P. Kudritzski for many useful 
discussions and insights concerning stellar atmospheres and non-LTE 
effects in giant stars. We also would like to thank an anonymous 
referee for several suggestions concerning the content and the 
cut of an early version of the current manuscript.
This work was partially supported by PRIN--INAF 2011 "Tracing the
formation and evolution of the Galactic halo with VST" (P.I.: M. Marconi)
and by PRIN--MIUR (2010LY5N2T) "Chemical and dynamical evolution of
the Milky Way and Local Group galaxies" (P.I.: F. Matteucci).
One of us (G.B.) thanks The Carnegie Observatories visitor programme for 
support as science visitor.
This publication makes use of data products from the Two Micron All Sky Survey,
which is a joint project of the University of Massachusetts and the Infrared
Processing and Analysis Center/California Institute of Technology, funded by
the National Aeronautics and Space Administration and the National Science
Foundation. 

%_____________________________________________________________________________________________________
\bibliographystyle{apj}

\clearpage 
%===================================================================================
\input{table_tot2_2_141111}
%===================================================================================

%+++++++++++++++++++++++++++++++++++++++++++++++++++++++++++++++++++++++++++++++++++
%			Figures 
%+++++++++++++++++++++++++++++++++++++++++++++++++++++++++++++++++++++++++++++++++++

\clearpage
%%%%%%%%%%%%% Fig 1 %%%%%%%%%%%%%%%%
\clearpage

%%%%%%%%%%%%% Fig 2 %%%%%%%%%%%%%%%%

%%%%%%%%%%%%% Fig 3 %%%%%%%%%%%%%%%%

%%%%%%%%%%%%% Fig 5 %%%%%%%%%%%%%%%%
\clearpage

%%%%%%%%%%%%% Fig 5 %%%%%%%%%%%%%%%%
\clearpage

%%%%%%%%%%%%% Fig 8 %%%%%%%%%%%%%%%%
\clearpage

%%%%%%%%%%%%% Fig 11 %%%%%%%%%%%%%%%%
\clearpage

%%%%%%%%%%%%% Fig 11 %%%%%%%%%%%%%%%%
\clearpage

%%%%%%%%%%%%% Fig 1 %%%%%%%%%%%%%%%%
\clearpage

%%%%%%%%%%%%% Fig 12 %%%%%%%%%%%%%%%%
\clearpage

%%%%%%%%%%%%% Fig 13 %%%%%%%%%%%%%%%%
\clearpage

%%%%%%%%%%%%% Fig 13 %%%%%%%%%%%%%%%%
\clearpage

%%%%%%%%%%%%%%%%%%%%%%%%%%%%%%%%%%%%%%%%%%%%%%%%%%%%%%%%%%%%%%
\end{document}

%% file: table_tot2_2_141111.tex
%++++++++++++++++++++++++++++++++++++++++++++++++++++++++++++++++++++++++++++++++++++++++++
%                               TABLE 4
%++++++++++++++++++++++++++++++++++++++++++++++++++++++++++++++++++++++++++++++++++++++++++

\scriptsize
\begin{deluxetable}{cc ccc}
\tablewidth{0pt}
\tabletypesize{\footnotesize}
\tablecaption{True distance moduli and reddenings for M4 available in the literature.\label{tab:table_dis}}
\tablehead{
$\mu$\tablenotemark{a} & $R_V$\tablenotemark{b} & $E(\bmv)$\tablenotemark{c}& Ref.\tablenotemark{d}  & Notes\tablenotemark{e} \\
                   mag &                   mag  &                   mag  &                        &                        \\ 
}
\startdata
11.28$\pm$0.06  & 3.62$\pm$0.07 & 0.37$\pm$0.01  & H12 & (1)  \\
11.18$\pm$0.18  & \ldots        & \ldots         & P95 & (2)  \\
11.19$\pm$0.01  & 3.8           & 0.34$\pm$0.03  & LJ  & (3)  \\
11.22$\pm$0.11  & 4             & 0.37$\pm$0.01  & DL  & (4)  \\
11.28$\pm$0.06  & 4             & 0.37           & L90 & (5)  \\
11.48           & 3.8           & 0.32           & B09 & (6)  \\
11.18$\pm$0.18  & 3.8           & 0.35$\pm$0.01  & H04 & (7)  \\
11.30$\pm$0.05  & 3.62$\pm$0.07 & 0.399$\pm$0.010& K13 & (8)  \\
11.37$\pm$0.08  & \ldots        & \ldots         & B03 & (9)  \\
\enddata
\tablenotetext{a}{True distance modulus and its error when estimated by the authors.}
\tablenotetext{b}{The ratio between absolute and selective extinction.}
\tablenotetext{c}{Mean reddening.}
\tablenotetext{d}{References: H12, \citet{hendricks2012}; P95: \citet{peterson1995};
LJ: \citet{liujanes1990b}; DL: \citet{dixonlong1993}; L90: \citet{longmore1990};
B09: \citet{bedin2009}; H04: \citet{hansen2004}; K13: \citet{kaluzny2013};
B03: \citet{bono2003}.}
\tablenotetext{e}{
(1) H12 derived a new reddening law for M4 by using both optical and NIR photometry.
The true distance modulus was estimated using the Zero Age Horizontal Branch (ZAHB).
(2) Astrometric distance based on proper motions and radial velocities. This distance modulus
is independent of reddening uncertainties.
(3) Baade--Wesselink distance based on optical Near Infrared (NIR) photometry of four RR Lyrae (V2, V15, V32, V33).
The individual reddening values are listed in their Table~5. They also assume an iron abundance
of [Fe/H]=--1.3$\pm$0.2.
(4) The distance is based on a new estimate of the reddening to M4, on a new metallicity
([Fe/H]=--1.10$\pm$0.25) and on distance estimates available in the literature.
(5) The distance is based on the $K$-band Period-Luminosity (PLK) relation of RR Lyrae stars. The error
on the distance is the error on the zero-point of the PLK relation of M4. The mean reddening
and the mean metallicity ([Fe/H]=--1.28) are from \citet{buonanno1989}.
(6) The distance is based on the fit between HB stars and the ZAHB. They provide an apparent
distance modulus in (m-M)$_{F606W}$ together with the extinction in the $V$ band ($A_V$=1.2).
They also assumed [Fe/H]=--1.07$\pm$0.01 ([$\alpha$/Fe]=0.39$\pm$0.05) by \citet{marino2008}.
(7) The distance is based on different distance estimates available in the literature and in
particular on the main sequence fitting provided by \citet{richer1997}.
(8) The distance is based on three eclipsing binaries. The dust-type parameter was adopted by
H12, the individual reddenings are listed in their Table~6.
(9) Distance based on the four RR Lyrae observed by \citet{liujanes1990b} and using a theoretical 
$K$-band Period-Luminosity-Metallicity relation provided by \citep[ see their Table~6]{bono2003}.  
}
\end{deluxetable}

%++++++++++++++++++++++++++++++++++++++++++++++++++++++++++++++++++++++++++++++++++++++++++
%                               TABLE 3
%++++++++++++++++++++++++++++++++++++++++++++++++++++++++++++++++++++++++++++++++++++++++++
\begin{deluxetable}{l rrc rrc rrc}
\tablewidth{0pt}
\tabletypesize{\footnotesize}
\tablecaption{Observed optical and NIR Period--Luminosity relations for RR Lyrae in M4.}
\tablehead{
Mag\tablenotemark{a} & a\tablenotemark{b}& b\tablenotemark{b}& $\sigma$\tablenotemark{b}& a\tablenotemark{c}& b\tablenotemark{c}& $\sigma$\tablenotemark{c}& a\tablenotemark{d} & b\tablenotemark{d} & $\sigma$\tablenotemark{d} \\
                       & mag & mag & mag & mag & mag  & mag  & mag & mag & mag  \\
}
\startdata
          & \multicolumn{3}{c}{FO} & \multicolumn{3}{c}{FU} & \multicolumn{3}{c}{FU$+$FO} \\
$R$ & 12.228 & -1.260 & 0.084 & 12.456 & -1.472 & 0.099 & 12.604 & -0.847 & 0.103 \\
 & $\pm$0.230 & $\pm$0.420 &  & $\pm$0.085 & $\pm$0.313 &  & $\pm$0.057 & $\pm$0.177 &  \\

$I$ & 11.609 & -1.549 & 0.072 & 11.858 & -1.724 & 0.070 & 12.004 & -1.137 & 0.075 \\
 & $\pm$0.195 & $\pm$0.356 &  & $\pm$0.085 & $\pm$0.311 &  & $\pm$0.047 & $\pm$0.144 &  \\

$J$ & 10.634 & -2.020 & 0.056 & 10.946 & -2.030 & 0.065 & 11.002 & -1.793 & 0.064 \\
 & $\pm$0.148 & $\pm$0.273 &  & $\pm$0.056 & $\pm$0.204 &  & $\pm$0.035 & $\pm$0.109 &  \\

$H$ & 10.232 & -2.340 & 0.037 & 10.537 & -2.215 & 0.050 & 10.492 & -2.408 & 0.046 \\
 & $\pm$0.097 & $\pm$0.179 &  & $\pm$0.047 & $\pm$0.176 &  & $\pm$0.027 & $\pm$0.082 &  \\

$K$ & 10.058 & -2.440 & 0.041 & 10.410 & -2.372 & 0.045 & 10.420 & -2.326 & 0.043 \\
 & $\pm$0.108 & $\pm$0.198 &  & $\pm$0.039 & $\pm$0.142 &  & $\pm$0.024 & $\pm$0.074 &  \\

\enddata
%\tablecomments{Table~2 is available in its entirety via the link to the machine-readable}
\tablenotetext{a}{PL relations of the form: $M_X$=a + b$\times$$\log\ P$.}
\tablenotetext{b}{Zero--point (a), slope (b) and standard deviation ($\sigma$) for first overtone (FO) pulsators.
The errors on the zero--point and on the slope are listed in the 2nd row.}
\tablenotetext{c}{Zero--point (a), slope (b) and standard deviation ($\sigma$) for fundamental (FU) pulsators.
The errors on the zero--point and on the slope are listed in the 2nd row.}
\tablenotetext{d}{Zero--point (a), slope (b) and standard deviation ($\sigma$) for 
for the entire sample (FU$+$FO) of RR Lyrae. The periods of FO variables were 
fundamentalized by adopting the following relation: $\log P_F$=$\log P_{FO}$+0.127. 
The errors on the zero--point and on the slope are listed in the 2nd row. 
}
\label{tab:table_pl}
\end{deluxetable}

%++++++++++++++++++++++++++++++++++++++++++++++++++++++++++++++++++++++++++++++++++++++++++
%                               TABLE 3
%++++++++++++++++++++++++++++++++++++++++++++++++++++++++++++++++++++++++++++++++++++++++++
\begin{deluxetable}{l r rrc rrc rrc}
\tablewidth{0pt}
\tabletypesize{\footnotesize}
\tablecaption{Observed optical Period--Wesenheit relations for RR Lyrae in M4.}
\tablehead{
PW\tablenotemark{a} & x\tablenotemark{b} & a\tablenotemark{c}& b\tablenotemark{c}& $\sigma$\tablenotemark{c}& a\tablenotemark{d}& b\tablenotemark{d}& $\sigma$\tablenotemark{d}& a\tablenotemark{e} & b\tablenotemark{e} & $\sigma$\tablenotemark{e} \\
                    &   & mag & mag  & mag  & mag & mag & mag  & mag & mag & mag \\
}
\startdata
          & \multicolumn{3}{c}{FO} & \multicolumn{3}{c}{FU} & \multicolumn{3}{c}{FU$+$FO} \\

$V$,\bmv  & 3.76  &  9.68     & -2.86     & 0.05 &  9.89     & -3.27     & 0.10 &  9.93     & -3.13     & 0.09 \\
       &       & $\pm$0.14 & $\pm$0.25 &       & $\pm$0.09 & $\pm$0.33 &       & $\pm$0.05 & $\pm$0.16 & \\
$R$,\bmr  & 1.91  &  9.53     & -2.89     & 0.06 &  9.72     & -3.40     & 0.07 &  9.78     & -3.18     & 0.07 \\
       &       & $\pm$0.17 & $\pm$0.30 &       & $\pm$0.07 & $\pm$0.25 &       & $\pm$0.04 & $\pm$0.12 & \\
$R$,\vmr  & 4.92  &  9.39     & -2.89     & 0.08 &  9.61     & -3.22     & 0.07 &  9.61     & -3.23     & 0.07 \\
       &       & $\pm$0.21 & $\pm$0.38 &       & $\pm$0.06 & $\pm$0.23 &       & $\pm$0.04 & $\pm$0.12 & \\
$I$,\bmi  & 0.92  &  9.73     & -2.60     & 0.05 &  9.98     & -2.93     & 0.08 & 10.05     & -2.67     & 0.07 \\
       &       & $\pm$0.13 & $\pm$0.24 &       & $\pm$0.07 & $\pm$0.26 &       & $\pm$0.04 & $\pm$0.12 & \\
$I$,\vmi  & 1.55  &  9.76     & -2.51     & 0.06 & 10.02     & -2.72     & 0.06 & 10.06     & -2.56     & 0.06 \\
       &       & $\pm$0.18 & $\pm$0.32 &       & $\pm$0.05 & $\pm$0.20 &       & $\pm$0.03 & $\pm$0.11 & \\
$I$,\rmi  & 2.73  &  9.83     & -2.47     & 0.09 & 10.19     & -2.48     & 0.08 & 10.25     & -2.25     & 0.08 \\
       &       & $\pm$0.24 & $\pm$0.44 &       & $\pm$0.07 & $\pm$0.28 &       & $\pm$0.05 & $\pm$0.15 & \\

\enddata
\tablenotetext{a}{PW relations of the form: $W(M_1,M_2-M_3)$=a + b$\times$$\log\ P$.}
\tablenotetext{b}{Color coefficient in Wesenheit magnitude: $x_{W(M_1,M_2-M_3)}$=$\frac{1}{A_{M_2}/A_{M_1}-A_{M_3}/A_{M_2}}$}
\tablenotetext{c}{Zero--point (a), slope (b) and standard deviation ($\sigma$) for FO variables.
The errors on the zero--point and on the slope are listed in the 2nd row.}
\tablenotetext{d}{Zero--point (a), slope (b) and standard deviation ($\sigma$) for FU variables.
The errors on the zero--point and on the slope are listed in the 2nd row.}
\tablenotetext{e}{Zero--point (a), slope (b) and standard deviation ($\sigma$) 
for the entire sample (FU$+$FO) of RR Lyrae. The periods of FO variables were 
fundamentalized by adopting the following relation: $\log P_F$=$\log P_{FO}$+0.127. 
The errors on the zero--point and on the slope are listed in the 2nd row.}
\label{tab:table_plw_opt}
\end{deluxetable}

%++++++++++++++++++++++++++++++++++++++++++++++++++++++++++++++++++++++++++++++++++++++++++
%                               TABLE 3
%++++++++++++++++++++++++++++++++++++++++++++++++++++++++++++++++++++++++++++++++++++++++++
\begin{deluxetable}{l r rrc rrc rrc}
\tablewidth{0pt}
\tabletypesize{\footnotesize}
\tablecaption{Observed optical--NIR Period--Wesenheit relations for RR Lyrae in M4.}
\tablehead{
PW\tablenotemark{a} & x\tablenotemark{b} & a\tablenotemark{c}& b\tablenotemark{c}& $\sigma$\tablenotemark{c}& a\tablenotemark{d}& b\tablenotemark{d}& $\sigma$\tablenotemark{d}& a\tablenotemark{e} & b\tablenotemark{e} & $\sigma$\tablenotemark{e} \\
                    &   & mag & mag  & mag  & mag & mag & mag  & mag & mag & mag \\
}
\startdata
          & \multicolumn{3}{c}{FO} & \multicolumn{3}{c}{FU} & \multicolumn{3}{c}{FU$+$FO} \\

$J$,\bmj  & 0.31 &  9.74     & -2.44     & 0.06 & 10.03     & -2.48     & 0.08 & 10.02     & -2.50     & 0.07 \\
       &      & $\pm$0.17 & $\pm$0.31 &       & $\pm$0.07 & $\pm$0.24 &       & $\pm$0.04 & $\pm$0.12 & \\
$J$,\vmj  & 0.43 &  9.75     & -2.41     & 0.07 & 10.04     & -2.43     & 0.09 & 10.03     & -2.46     & 0.08 \\
       &      & $\pm$0.18 & $\pm$0.33 &       & $\pm$0.08 & $\pm$0.28 &       & $\pm$0.04 & $\pm$0.14 & \\
$J$,\rmj  & 0.57 &  9.68     & -2.53     & 0.06 & 10.07     & -2.43     & 0.08 & 10.09     & -2.33     & 0.08 \\
       &      & $\pm$0.16 & $\pm$0.30 &       & $\pm$0.07 & $\pm$0.27 &       & $\pm$0.04 & $\pm$0.13 & \\
$J$,\imj  & 0.99 &  9.62     & -2.58     & 0.08 & 10.03     & -2.53     & 0.10 & 10.09     & -2.27     & 0.09 \\
       &      & $\pm$0.22 & $\pm$0.40 &       & $\pm$0.08 & $\pm$0.31 &       & $\pm$0.05 & $\pm$0.16 & \\
$H$,\bmh  & 0.18 &  9.65     & -2.64     & 0.03 &  9.94     & -2.51     & 0.04 &  9.84     & -2.94     & 0.05 \\
       &      & $\pm$0.08 & $\pm$0.15 &       & $\pm$0.04 & $\pm$0.13 &       & $\pm$0.03 & $\pm$0.08 & \\
$H$,\vmh  & 0.24 &  9.65     & -2.63     & 0.03 &  9.94     & -2.48     & 0.04 &  9.84     & -2.93     & 0.05 \\
       &      & $\pm$0.09 & $\pm$0.16 &       & $\pm$0.04 & $\pm$0.14 &       & $\pm$0.03 & $\pm$0.09 & \\
$H$,\rmh  & 0.30 &  9.68     & -2.60     & 0.03 &  9.96     & -2.45     & 0.04 &  9.85     & -2.91     & 0.05 \\
       &      & $\pm$0.09 & $\pm$0.17 &       & $\pm$0.04 & $\pm$0.15 &       & $\pm$0.03 & $\pm$0.09 & \\
$H$,\imh  & 0.46 &  9.64     & -2.63     & 0.05 &  9.93     & -2.45     & 0.05 &  9.80     & -2.98     & 0.06 \\
       &      & $\pm$0.13 & $\pm$0.24 &       & $\pm$0.05 & $\pm$0.19 &       & $\pm$0.04 & $\pm$0.11 & \\
$K$,\bmk  & 0.11 &  9.69     & -2.63     & 0.04 & 10.04     & -2.56     & 0.05 & 10.02     & -2.63     & 0.05 \\
       &      & $\pm$0.10 & $\pm$0.19 &       & $\pm$0.04 & $\pm$0.15 &       & $\pm$0.02 & $\pm$0.08 & \\
$K$,\vmk  & 0.14 &  9.69     & -2.62     & 0.04 & 10.04     & -2.55     & 0.05 & 10.02     & -2.62     & 0.05 \\
       &      & $\pm$0.10 & $\pm$0.19 &       & $\pm$0.04 & $\pm$0.16 &       & $\pm$0.03 & $\pm$0.08 & \\
$K$,\rmk  & 0.17 &  9.66     & -2.67     & 0.04 & 10.05     & -2.55     & 0.05 & 10.04     & -2.59     & 0.05 \\
       &      & $\pm$0.10 & $\pm$0.19 &       & $\pm$0.04 & $\pm$0.16 &       & $\pm$0.03 & $\pm$0.08 & \\
$K$,\imk  & 0.25 &  9.65     & -2.70     & 0.04 & 10.04     & -2.53     & 0.05 & 10.02     & -2.61     & 0.05 \\
       &      & $\pm$0.11 & $\pm$0.21 &       & $\pm$0.04 & $\pm$0.16 &       & $\pm$0.03 & $\pm$0.08 & \\
$K$,\jmk  & 0.69 &  9.66     & -2.73     & 0.03 & 10.04     & -2.61     & 0.08 & 10.02     & -2.69     & 0.07 \\
       &      & $\pm$0.09 & $\pm$0.17 &       & $\pm$0.07 & $\pm$0.25 &       & $\pm$0.04 & $\pm$0.12 & \\
\multicolumn{10}{c}{--- Three-bands ---} \\
$J$,\bmi  & 0.46 &  9.67     & -2.59     & 0.05 & 10.03     & -2.59     & 0.04 & 10.05     & -2.49     & 0.04 \\
       &      & $\pm$0.13 & $\pm$0.23 &       & $\pm$0.04 & $\pm$0.13 &       & $\pm$0.02 & $\pm$0.07 & \\
$J$,\bmr  & 0.69 &  9.63     & -2.66     & 0.05 & 10.00     & -2.56     & 0.07 &  9.99     & -2.60     & 0.06 \\
       &      & $\pm$0.14 & $\pm$0.26 &       & $\pm$0.06 & $\pm$0.21 &       & $\pm$0.03 & $\pm$0.10 & \\
$J$,\vmi  & 0.77 &  9.69     & -2.54     & 0.05 & 10.03     & -2.56     & 0.04 & 10.06     & -2.44     & 0.04 \\
       &      & $\pm$0.13 & $\pm$0.24 &       & $\pm$0.04 & $\pm$0.14 &       & $\pm$0.02 & $\pm$0.08 & \\
$H$,\bmi  & 0.29 &  9.67     & -2.62     & 0.03 &  9.95     & -2.55     & 0.04 &  9.87     & -2.90     & 0.04 \\
       &      & $\pm$0.07 & $\pm$0.13 &       & $\pm$0.04 & $\pm$0.13 &       & $\pm$0.02 & $\pm$0.07 & \\
$H$,\bmr  & 0.44 &  9.64     & -2.67     & 0.03 &  9.92     & -2.61     & 0.04 &  9.84     & -2.96     & 0.05 \\
       &      & $\pm$0.09 & $\pm$0.17 &       & $\pm$0.04 & $\pm$0.15 &       & $\pm$0.03 & $\pm$0.08 & \\
$H$,\vmi  & 0.49 &  9.68     & -2.59     & 0.02 &  9.96     & -2.50     & 0.04 &  9.87     & -2.87     & 0.04 \\
       &      & $\pm$0.06 & $\pm$0.11 &       & $\pm$0.04 & $\pm$0.14 &       & $\pm$0.03 & $\pm$0.08 & \\
$K$,\bmi  & 0.19 &  9.66     & -2.68     & 0.04 & 10.03     & -2.58     & 0.05 & 10.02     & -2.63     & 0.04 \\
       &      & $\pm$0.10 & $\pm$0.18 &       & $\pm$0.04 & $\pm$0.15 &       & $\pm$0.02 & $\pm$0.08 & \\
$K$,\bmr  & 0.28 &  9.64     & -2.71     & 0.04 & 10.02     & -2.58     & 0.05 & 10.00     & -2.66     & 0.05 \\
       &      & $\pm$0.10 & $\pm$0.19 &       & $\pm$0.04 & $\pm$0.16 &       & $\pm$0.03 & $\pm$0.08 & \\
$K$,\vmi  & 0.31 &  9.67     & -2.66     & 0.04 & 10.04     & -2.57     & 0.05 & 10.03     & -2.61     & 0.05 \\
       &      & $\pm$0.10 & $\pm$0.18 &       & $\pm$0.04 & $\pm$0.16 &       & $\pm$0.03 & $\pm$0.08 & \\

\enddata
\tablenotetext{a}{PW relations of the form: $W(M_1,M_2-M_3)$=a + b$\times$$\log\ P$. 
$M_3$$\neq$$M_1$ only for three-band Wesenheit magnitudes}
\tablenotetext{b}{Color coefficient in Wesenheit magnitude: $x_{W(M_1,M_2-M_3)}$=$\frac{1}{A_{M_2}/A_{M_1}-A_{M_3}/A_{M_2}}$}
\tablenotetext{c}{Zero--point (a), slope (b) and standard deviation ($\sigma$) for FO pulsators.
The errors on zero--point and slope are listed in the 2nd row.}
\tablenotetext{d}{Zero--point (a), slope (b) and standard deviation ($\sigma$) for FU pulsators.
The errors on zero--point and slope are listed in the 2nd row.}
\tablenotetext{e}{Zero--point (a), slope (b) and standard deviation ($\sigma$) for 
for the entire sample (FU$+$FO) of RR Lyrae. The periods of FO variables were 
fundamentalized by adopting the following relation: $\log P_F$=$\log P_{FO}$+0.127. 
The errors on zero--point and slope are listed in the 2nd row.}
\label{tab:table_plw_nir}
\end{deluxetable}

%++++++++++++++++++++++++++++++++++++++++++++++++++++++++++++++++++++++++++++++++++++++++++
%                               TABLE 3
%++++++++++++++++++++++++++++++++++++++++++++++++++++++++++++++++++++++++++++++++++++++++++
\begin{deluxetable}{l r rrrc rrrc rrrc}
\tablewidth{0pt}
\tabletypesize{\footnotesize}
\tablecaption{Theoretical optical and NIR Period--Wesenheit--Metallicity relations for RR Lyrae in M4.}
\tablehead{
PWZ\tablenotemark{a} & x\tablenotemark{b} & a\tablenotemark{c}& b\tablenotemark{c}& c\tablenotemark{c}& $\sigma$\tablenotemark{c}& a\tablenotemark{d}& b\tablenotemark{d}& c\tablenotemark{d}& $\sigma$\tablenotemark{d}& a\tablenotemark{e} & b\tablenotemark{e} & c\tablenotemark{e}& $\sigma$\tablenotemark{e} \\
                    &   & mag & mag & mag & mag & mag  & mag  & mag & mag & mag  & mag & mag & mag \\
}
\startdata
          & \multicolumn{4}{c}{FO} & \multicolumn{4}{c}{FU} & \multicolumn{4}{c}{FU$+$FO} \\
$V$,\bmv & 3.76  & -1.800 & -2.858 & -0.007 & 0.049 & -1.487 & -3.031 & -0.066 & 0.106 & -1.444 & -2.848 & -0.043 & 0.102 \\
      &       & $\pm$0.055 & $\pm$0.108 & $\pm$0.009 & & $\pm$0.022 & $\pm$0.064 & $\pm$0.011 & & $\pm$0.019 & $\pm$0.049 & $\pm$0.009 \\
$R$,\bmr & 1.91  & -1.793 & -2.936 &  0.036 & 0.040 & -1.389 & -2.922 &  0.010 & 0.082 & -1.365 & -2.800 &  0.023 & 0.078 \\
      &       & $\pm$0.044 & $\pm$0.088 & $\pm$0.007 & & $\pm$0.017 & $\pm$0.050 & $\pm$0.009 & & $\pm$0.014 & $\pm$0.038 & $\pm$0.007 \\
$R$,\vmr & 4.92  & -1.787 & -3.016 &  0.080 & 0.031 & -1.288 & -2.810 &  0.088 & 0.060 & -1.283 & -2.750 &  0.090 & 0.056 \\
      &       & $\pm$0.035 & $\pm$0.068 & $\pm$0.006 & & $\pm$0.013 & $\pm$0.036 & $\pm$0.006 & & $\pm$0.010 & $\pm$0.027 & $\pm$0.005 \\
$I$,\bmi & 0.92  & -1.639 & -2.878 &  0.094 & 0.026 & -1.149 & -2.648 &  0.095 & 0.048 & -1.139 & -2.568 &  0.101 & 0.048 \\
      &       & $\pm$0.029 & $\pm$0.057 & $\pm$0.005 & & $\pm$0.010 & $\pm$0.029 & $\pm$0.005 & & $\pm$0.009 & $\pm$0.023 & $\pm$0.004 \\
$I$,\vmi & 1.55  & -1.586 & -2.884 &  0.127 & 0.019 & -1.039 & -2.524 &  0.147 & 0.034 & -1.039 & -2.476 &  0.147 & 0.036 \\
      &       & $\pm$0.021 & $\pm$0.041 & $\pm$0.003 & & $\pm$0.007 & $\pm$0.021 & $\pm$0.004 & & $\pm$0.007 & $\pm$0.017 & $\pm$0.003 \\
$I$,\rmi & 2.73  & -1.494 & -2.824 &  0.148 & 0.016 & -0.924 & -2.392 &  0.175 & 0.031 & -0.927 & -2.350 &  0.174 & 0.034 \\
      &       & $\pm$0.018 & $\pm$0.036 & $\pm$0.003 & & $\pm$0.006 & $\pm$0.018 & $\pm$0.003 & & $\pm$0.006 & $\pm$0.016 & $\pm$0.003 \\
$J$,\bmj & 0.31  & -1.536 & -2.776 &  0.134 & 0.020 & -1.005 & -2.417 &  0.152 & 0.031 & -1.005 & -2.367 &  0.153 & 0.034 \\
      &       & $\pm$0.023 & $\pm$0.044 & $\pm$0.004 & & $\pm$0.006 & $\pm$0.019 & $\pm$0.003 & & $\pm$0.006 & $\pm$0.016 & $\pm$0.003 \\
$J$,\vmj & 0.43  & -1.512 & -2.769 &  0.146 & 0.019 & -0.961 & -2.362 &  0.172 & 0.030 & -0.965 & -2.323 &  0.171 & 0.033 \\
      &       & $\pm$0.021 & $\pm$0.041 & $\pm$0.003 & & $\pm$0.006 & $\pm$0.018 & $\pm$0.003 & & $\pm$0.006 & $\pm$0.016 & $\pm$0.003 \\
$J$,\rmj & 0.57  & -1.486 & -2.745 &  0.153 & 0.019 & -0.930 & -2.318 &  0.180 & 0.032 & -0.934 & -2.282 &  0.178 & 0.035 \\
      &       & $\pm$0.022 & $\pm$0.043 & $\pm$0.004 & & $\pm$0.007 & $\pm$0.019 & $\pm$0.003 & & $\pm$0.006 & $\pm$0.017 & $\pm$0.003 \\
$J$,\imj & 0.99  & -1.484 & -2.724 &  0.154 & 0.021 & -0.932 & -2.299 &  0.181 & 0.033 & -0.936 & -2.264 &  0.180 & 0.036 \\
      &       & $\pm$0.023 & $\pm$0.045 & $\pm$0.004 & & $\pm$0.007 & $\pm$0.020 & $\pm$0.004 & & $\pm$0.007 & $\pm$0.017 & $\pm$0.003 \\
$H$,\bmh & 0.18  & -1.621 & -2.916 &  0.148 & 0.015 & -1.083 & -2.532 &  0.175 & 0.029 & -1.092 & -2.541 &  0.171 & 0.029 \\
      &       & $\pm$0.016 & $\pm$0.032 & $\pm$0.003 & & $\pm$0.006 & $\pm$0.018 & $\pm$0.003 & & $\pm$0.005 & $\pm$0.014 & $\pm$0.003 \\
$H$,\vmh & 0.24  & -1.612 & -2.919 &  0.156 & 0.013 & -1.062 & -2.508 &  0.187 & 0.027 & -1.075 & -2.526 &  0.182 & 0.028 \\
      &       & $\pm$0.015 & $\pm$0.029 & $\pm$0.002 & & $\pm$0.006 & $\pm$0.017 & $\pm$0.003 & & $\pm$0.005 & $\pm$0.014 & $\pm$0.003 \\
$H$,\rmh & 0.30  & -1.603 & -2.914 &  0.160 & 0.013 & -1.051 & -2.492 &  0.192 & 0.027 & -1.064 & -2.515 &  0.186 & 0.028 \\
      &       & $\pm$0.014 & $\pm$0.028 & $\pm$0.002 & & $\pm$0.006 & $\pm$0.016 & $\pm$0.003 & & $\pm$0.005 & $\pm$0.013 & $\pm$0.002 \\
$H$,\imh & 0.46  & -1.616 & -2.925 &  0.161 & 0.013 & -1.067 & -2.505 &  0.195 & 0.027 & -1.081 & -2.535 &  0.188 & 0.028 \\
      &       & $\pm$0.014 & $\pm$0.027 & $\pm$0.002 & & $\pm$0.006 & $\pm$0.016 & $\pm$0.003 & & $\pm$0.005 & $\pm$0.013 & $\pm$0.002 \\
$K$,\bmk & 0.11  & -1.564 & -2.852 &  0.149 & 0.016 & -1.023 & -2.455 &  0.177 & 0.027 & -1.031 & -2.453 &  0.173 & 0.028 \\
      &       & $\pm$0.017 & $\pm$0.034 & $\pm$0.003 & & $\pm$0.006 & $\pm$0.016 & $\pm$0.003 & & $\pm$0.005 & $\pm$0.014 & $\pm$0.003 \\
$K$,\vmk & 0.14  & -1.557 & -2.851 &  0.154 & 0.015 & -1.009 & -2.438 &  0.184 & 0.027 & -1.019 & -2.441 &  0.179 & 0.028 \\
      &       & $\pm$0.016 & $\pm$0.032 & $\pm$0.003 & & $\pm$0.006 & $\pm$0.016 & $\pm$0.003 & & $\pm$0.005 & $\pm$0.014 & $\pm$0.003 \\
$K$,\rmk & 0.17  & -1.550 & -2.847 &  0.156 & 0.015 & -1.001 & -2.427 &  0.187 & 0.027 & -1.011 & -2.432 &  0.182 & 0.028 \\
      &       & $\pm$0.016 & $\pm$0.032 & $\pm$0.003 & & $\pm$0.006 & $\pm$0.016 & $\pm$0.003 & & $\pm$0.005 & $\pm$0.014 & $\pm$0.003 \\
$K$,\imk & 0.25  & -1.554 & -2.848 &  0.156 & 0.015 & -1.006 & -2.429 &  0.188 & 0.027 & -1.017 & -2.438 &  0.183 & 0.028 \\
      &       & $\pm$0.016 & $\pm$0.032 & $\pm$0.003 & & $\pm$0.006 & $\pm$0.016 & $\pm$0.003 & & $\pm$0.005 & $\pm$0.013 & $\pm$0.002 \\
$K$,\jmk & 0.69  & -1.579 & -2.891 &  0.157 & 0.014 & -1.032 & -2.475 &  0.190 & 0.026 & -1.045 & -2.498 &  0.184 & 0.027 \\
      &       & $\pm$0.015 & $\pm$0.030 & $\pm$0.002 & & $\pm$0.005 & $\pm$0.016 & $\pm$0.003 & & $\pm$0.005 & $\pm$0.013 & $\pm$0.002 \\
$H$,\jmh & 1.72  & -1.731 & -3.099 &  0.167 & 0.018 & -1.184 & -2.683 &  0.206 & 0.037 & -1.206 & -2.770 &  0.195 & 0.038 \\
      &       & $\pm$0.020 & $\pm$0.040 & $\pm$0.003 & & $\pm$0.008 & $\pm$0.022 & $\pm$0.004 & & $\pm$0.007 & $\pm$0.018 & $\pm$0.003 \\
$K$,\hmk & 1.81  & -1.477 & -2.753 &  0.150 & 0.020 & -0.931 & -2.336 &  0.179 & 0.031 & -0.938 & -2.317 &  0.176 & 0.032 \\
      &       & $\pm$0.022 & $\pm$0.043 & $\pm$0.004 & & $\pm$0.006 & $\pm$0.019 & $\pm$0.003 & & $\pm$0.006 & $\pm$0.016 & $\pm$0.003 \\
\multicolumn{13}{c}{--- Three-bands ---} \\
$J$,\bmr & 0.46  & -1.598 & -2.814 &  0.110 & 0.023 & -1.097 & -2.538 &  0.118 & 0.037 & -1.091 & -2.470 &  0.122 & 0.039 \\
      &       & $\pm$0.026 & $\pm$0.051 & $\pm$0.004 & & $\pm$0.008 & $\pm$0.023 & $\pm$0.004 & & $\pm$0.007 & $\pm$0.019 & $\pm$0.004 \\
$J$,\bmi & 0.69  & -1.561 & -2.800 &  0.124 & 0.021 & -1.040 & -2.472 &  0.138 & 0.033 & -1.037 & -2.415 &  0.140 & 0.036 \\
      &       & $\pm$0.023 & $\pm$0.046 & $\pm$0.004 & & $\pm$0.007 & $\pm$0.020 & $\pm$0.003 & & $\pm$0.007 & $\pm$0.017 & $\pm$0.003 \\
$J$,\vmi & 0.77  & -1.535 & -2.804 &  0.140 & 0.018 & -0.985 & -2.410 &  0.165 & 0.030 & -0.987 & -2.370 &  0.164 & 0.033 \\
      &       & $\pm$0.020 & $\pm$0.040 & $\pm$0.003 & & $\pm$0.006 & $\pm$0.018 & $\pm$0.003 & & $\pm$0.006 & $\pm$0.016 & $\pm$0.003 \\
$H$,\bmr & 0.29  & -1.647 & -2.919 &  0.131 & 0.018 & -1.129 & -2.591 &  0.150 & 0.035 & -1.133 & -2.580 &  0.149 & 0.033 \\
      &       & $\pm$0.020 & $\pm$0.040 & $\pm$0.003 & & $\pm$0.007 & $\pm$0.021 & $\pm$0.004 & & $\pm$0.006 & $\pm$0.016 & $\pm$0.003 \\
$H$,\bmi & 0.44  & -1.623 & -2.910 &  0.140 & 0.016 & -1.093 & -2.550 &  0.163 & 0.031 & -1.099 & -2.545 &  0.161 & 0.031 \\
      &       & $\pm$0.018 & $\pm$0.036 & $\pm$0.003 & & $\pm$0.007 & $\pm$0.019 & $\pm$0.003 & & $\pm$0.006 & $\pm$0.015 & $\pm$0.003 \\
$H$,\vmi & 0.49  & -1.607 & -2.912 &  0.150 & 0.014 & -1.058 & -2.511 &  0.180 & 0.028 & -1.068 & -2.517 &  0.175 & 0.029 \\
      &       & $\pm$0.016 & $\pm$0.031 & $\pm$0.003 & & $\pm$0.006 & $\pm$0.017 & $\pm$0.003 & & $\pm$0.005 & $\pm$0.014 & $\pm$0.003 \\
$K$,\bmr & 0.19  & -1.586 & -2.860 &  0.138 & 0.017 & -1.059 & -2.500 &  0.160 & 0.029 & -1.064 & -2.487 &  0.158 & 0.030 \\
      &       & $\pm$0.019 & $\pm$0.038 & $\pm$0.003 & & $\pm$0.006 & $\pm$0.018 & $\pm$0.003 & & $\pm$0.006 & $\pm$0.014 & $\pm$0.003 \\
$K$,\bmi & 0.28  & -1.571 & -2.854 &  0.144 & 0.016 & -1.035 & -2.474 &  0.169 & 0.028 & -1.042 & -2.464 &  0.166 & 0.029 \\
      &       & $\pm$0.018 & $\pm$0.036 & $\pm$0.003 & & $\pm$0.006 & $\pm$0.017 & $\pm$0.003 & & $\pm$0.005 & $\pm$0.014 & $\pm$0.003 \\
$K$,\vmi & 0.31  & -1.561 & -2.855 &  0.150 & 0.015 & -1.013 & -2.448 &  0.179 & 0.027 & -1.022 & -2.446 &  0.175 & 0.028 \\
      &       & $\pm$0.017 & $\pm$0.033 & $\pm$0.003 & & $\pm$0.006 & $\pm$0.016 & $\pm$0.003 & & $\pm$0.005 & $\pm$0.014 & $\pm$0.003 \\
\enddata
\tablenotetext{a}{The PWZ relations of the form: 
$W(M_1,M_2-M_3)$= a + b$\times$$\log\ P$ + c$\times$$[Fe/H]$. $M_3$$\neq$$M_1$
only for three-band Wesenheit magnitudes}
\tablenotetext{b}{Color coefficient in Wesenheit magnitude: $x_{W(M_1,M_2-M_3)}$=$\frac{1}{A_{M_2}/A_{M_1}-A_{M_3}/A_{M_2}}$}
\tablenotetext{c}{Zero--point (a), slope (b), metallicity term (c) and standard deviation ($\sigma$) for FO pulsators.
The errors on the zero--point, slope and metallicity term are listed in the 2nd row.}
\tablenotetext{d}{Zero--point (a), slope (b), metallicity term (c) and standard deviation ($\sigma$) for FU pulsators.
The errors on the zero--point, slope and metallicity term are listed in the 2nd row.}
\tablenotetext{e}{Zero--point (a), slope (b), metallicity term (c) and standard deviation ($\sigma$) for 
for the entire sample (FU$+$FO) of RR Lyrae. The periods of FO variables were 
fundamentalized by adopting the following relation: $\log P_F$=$\log P_{FO}$+0.127. 
The errors on the zero--point, slope and metallicity term are listed in the 2nd row.}
\label{tab:table_plw_th_nir}
\end{deluxetable}

%++++++++++++++++++++++++++++++++++++++++++++++++++++++++++++++++++++++++++++++++++++++++++
%                               TABLE 5
%++++++++++++++++++++++++++++++++++++++++++++++++++++++++++++++++++++++++++++++++++++++++++
\scriptsize
\begin{deluxetable}{c ccccc c c c}
\tablewidth{0pt}
\tabletypesize{\footnotesize}
\tablecaption{RR Lyr photometric and physical properties from literature.}
\tablehead{
Period\tablenotemark{a} & $B$\tablenotemark{b}& $V$\tablenotemark{b}& $J$\tablenotemark{c}& $H$\tablenotemark{c}& $K$\tablenotemark{c}& $\pi$\tablenotemark{d} & [Fe/H]\tablenotemark{e}& $E(\bmv)$\tablenotemark{f} \\
days &     \multicolumn{5}{c}{mag} & \mas & & mag \\
}
\startdata
0.5668386$\pm$0.0000016 & 8.09$\pm$0.04  & 7.74$\pm$0.02  & 6.74$\pm$0.02 & 6.60$\pm$0.03 &  6.50$\pm$0.02 & 3.77$\pm$0.13 & -1.50$\pm$0.13 & 0.02$\pm$0.03 \\
\enddata
\tablenotetext{a}{Pulsation period from \citet{kolenberg2006}}
\tablenotetext{b}{Intensity averaged mean magnitude estimated with a spline fit to the {\it BV \/}
photoelectric photometry provided by \citet{szeidl1997}.}
\tablenotetext{c}{Intensity averaged mean {\it JHK \/} magnitude (2MASS, photometric system) 
provided by \citet{sollima2008}.}
\tablenotetext{d}{Trigonometric parallax from \citet{benedict2002}}
%\tablenotetext{e}{Iron abundance provided by \citet{kolenberg2010}, note that to provide an homogeneous
%metallicity scale, we took account for the difference in the solar iron abundance adopted
%by \citet{kolenberg2010} \citep[$\log\epsilon_{Fe}$=7.45, ][]{asplund2005} and by \citet{carr09}
%\citep[$\log\epsilon_{Fe}$=7.54, ][]{gratton2003} in defining the GC metallicity scale.}
\tablenotetext{e}{Iron abundance provided by \citet{kolenberg2010}, note that to provide an homogeneous
metallicity scale, we took account for the difference in the solar iron abundance adopted
by \citet[][$\log\epsilon_{Fe}$=7.45, \citet{asplund2005}]{kolenberg2010} and 
by \citet[][$\log\epsilon_{Fe}$=7.54, \citet{gratton2003}]{carr09}
in defining the GC metallicity scale.}
\tablenotetext{f}{Reddening according to \citet{sollima2008}.}
\label{tab:table_rrlyr}
\end{deluxetable}

%++++++++++++++++++++++++++++++++++++++++++++++++++++++++++++++++++++++++++++++++++++++++++
%                               TABLE 5
%++++++++++++++++++++++++++++++++++++++++++++++++++++++++++++++++++++++++++++++++++++++++++
\scriptsize
\begin{deluxetable}{l cc cc}
\tablewidth{0pt}
\tabletypesize{\footnotesize}
\tablecaption{True distance moduli based on observed Period--Luminosity and
Period--Wesenheit relations calibrated with RR Lyr.}
\tablehead{
PL-PW\tablenotemark{a} & $\mu$\tablenotemark{b}& $\sigma_{\mu}$\tablenotemark{c}& $\mu$\tablenotemark{b}& $\sigma_{\mu}$\tablenotemark{c} \\
             & \multicolumn{2}{c}{mag} & \multicolumn{2}{c}{mag} \\
}
\startdata
        &  \multicolumn{2}{c}{FU} & \multicolumn{2}{c}{FU$+$FO} \\
\multicolumn{5}{c}{--- PL ---} \\
     $J$  & 11.374$\pm$0.010 & 0.056 & 11.368$\pm$0.009 & 0.059 \\
     $H$  & 11.296$\pm$0.011 & 0.054 & 11.298$\pm$0.008 & 0.048 \\
     $K$  & 11.382$\pm$0.008 & 0.044 & 11.382$\pm$0.006 & 0.042 \\
\multicolumn{5}{c}{--- PW ---} \\
$V$,\bmv  & 11.166$\pm$0.019 & 0.103 & 11.159$\pm$0.014 & 0.090 \\
$J$,\bmj  & 11.333$\pm$0.012 & 0.067 & 11.329$\pm$0.009 & 0.063 \\
$J$,\vmj  & 11.342$\pm$0.013 & 0.073 & 11.337$\pm$0.010 & 0.068 \\
$H$,\bmh  & 11.245$\pm$0.009 & 0.043 & 11.251$\pm$0.008 & 0.047 \\
$H$,\vmh  & 11.246$\pm$0.010 & 0.047 & 11.252$\pm$0.008 & 0.050 \\
$K$,\bmk  & 11.362$\pm$0.008 & 0.047 & 11.364$\pm$0.007 & 0.044 \\
$K$,\vmk  & 11.364$\pm$0.008 & 0.047 & 11.367$\pm$0.007 & 0.044 \\
$K$,\jmk  & 11.378$\pm$0.013 & 0.073 & 11.383$\pm$0.009 & 0.063 \\

\enddata
\tablenotetext{a}{Adopted PL or PW relation.}
\tablenotetext{b}{Mean distance modulus and its error based on FU and FU$+$FO variables.}
\tablenotetext{c}{Standard deviation of the distance modulus based on FU and FU$+$FO 
variables.}
\label{tab:table_dmod_obs_pl}
\end{deluxetable}

%++++++++++++++++++++++++++++++++++++++++++++++++++++++++++++++++++++++++++++++++++++++++++
%                               TABLE 5
%++++++++++++++++++++++++++++++++++++++++++++++++++++++++++++++++++++++++++++++++++++++++++
\scriptsize
\begin{deluxetable}{l cc cc cc c}
\tablewidth{0pt}
\tabletypesize{\footnotesize}
\tablecaption{True distance moduli based on predicted Period--Luminosity--Metallicity relations.}
\tablehead{
PLZ\tablenotemark{a} & $\mu$\tablenotemark{b}& $\sigma_{\mu}$\tablenotemark{c}& $\mu$\tablenotemark{b}& $\sigma_{\mu}$\tablenotemark{c}& $\mu$\tablenotemark{b}& $\sigma_{\mu}$\tablenotemark{c} \\
             & \multicolumn{2}{c}{mag} & \multicolumn{2}{c}{mag} & \multicolumn{2}{c}{mag} \\
}
\startdata
        & \multicolumn{2}{c}{FO} & \multicolumn{2}{c}{FU} & \multicolumn{2}{c}{FU$+$FO} \\
    $R$  & 11.225$\pm$0.026 & 0.091 & 11.310$\pm$0.018 & 0.097 & 11.315$\pm$0.017 & 0.108 \\
    $I$  & 11.249$\pm$0.023 & 0.079 & 11.320$\pm$0.013 & 0.068 & 11.323$\pm$0.013 & 0.081 \\
    $J$  & 11.220$\pm$0.017 & 0.060 & 11.257$\pm$0.010 & 0.057 & 11.265$\pm$0.009 & 0.061 \\
    $H$  & 11.271$\pm$0.009 & 0.034 & 11.252$\pm$0.011 & 0.055 & 11.273$\pm$0.008 & 0.048 \\
    $K$  & 11.281$\pm$0.011 & 0.041 & 11.288$\pm$0.008 & 0.043 & 11.298$\pm$0.006 & 0.042 \\

\enddata
\tablenotetext{a}{Adopted PLZ relation.}
\tablenotetext{b}{Mean distance modulus and its error based on FO, FU and FU$+$FO variables.}
\tablenotetext{c}{Standard deviation of the distance modulus based on FO, FU and FU$+$FO 
variables.}
\label{tab:table_dmod_th_pl}
\end{deluxetable}

%++++++++++++++++++++++++++++++++++++++++++++++++++++++++++++++++++++++++++++++++++++++++++
%                               TABLE 5
%++++++++++++++++++++++++++++++++++++++++++++++++++++++++++++++++++++++++++++++++++++++++++
\scriptsize
\begin{deluxetable}{l cc cc cc}
\tablewidth{0pt}
\tabletypesize{\footnotesize}
\tablecaption{True distance moduli based on predicted optical and NIR 
Period--Wesenheit--Metallicity relations.}
\tablehead{
PWZ\tablenotemark{a} & $\mu$\tablenotemark{b}& $\sigma_{\mu}$\tablenotemark{c}& $\mu$\tablenotemark{b}& $\sigma_{\mu}$\tablenotemark{c}& $\mu$\tablenotemark{b}& $\sigma_{\mu}$\tablenotemark{c} \\
             & \multicolumn{2}{c}{mag} & \multicolumn{2}{c}{mag} & \multicolumn{2}{c}{mag} \\
}
\startdata
        & \multicolumn{2}{c}{FO} & \multicolumn{2}{c}{FU} & \multicolumn{2}{c}{FU$+$FO} \\
$V$,\bmv  & 11.472$\pm$0.014 & 0.050 & 11.372$\pm$0.019 & 0.104 & 11.415$\pm$0.014 & 0.093 \\
$R$,\bmr  & 11.338$\pm$0.017 & 0.058 & 11.248$\pm$0.015 & 0.077 & 11.287$\pm$0.013 & 0.078 \\
$R$,\vmr  & 11.201$\pm$0.021 & 0.073 & 11.103$\pm$0.014 & 0.070 & 11.144$\pm$0.013 & 0.082 \\
$I$,\bmi  & 11.326$\pm$0.014 & 0.048 & 11.312$\pm$0.015 & 0.080 & 11.329$\pm$0.011 & 0.072 \\
$I$,\vmi  & 11.279$\pm$0.019 & 0.066 & 11.275$\pm$0.012 & 0.063 & 11.289$\pm$0.010 & 0.062 \\
$I$,\rmi  & 11.314$\pm$0.033 & 0.116 & 11.335$\pm$0.016 & 0.081 & 11.343$\pm$0.015 & 0.091 \\
$J$,\bmj  & 11.232$\pm$0.013 & 0.048 & 11.234$\pm$0.011 & 0.063 & 11.247$\pm$0.009 & 0.060 \\
$J$,\vmj  & 11.209$\pm$0.014 & 0.051 & 11.223$\pm$0.012 & 0.067 & 11.232$\pm$0.010 & 0.063 \\
$J$,\rmj  & 11.209$\pm$0.015 & 0.053 & 11.238$\pm$0.013 & 0.072 & 11.242$\pm$0.011 & 0.068 \\
$J$,\imj  & 11.170$\pm$0.014 & 0.048 & 11.238$\pm$0.018 & 0.093 & 11.227$\pm$0.014 & 0.087 \\
$H$,\bmh  & 11.270$\pm$0.004 & 0.013 & 11.250$\pm$0.009 & 0.044 & 11.274$\pm$0.007 & 0.039 \\
$H$,\vmh  & 11.259$\pm$0.004 & 0.013 & 11.246$\pm$0.010 & 0.048 & 11.268$\pm$0.007 & 0.041 \\
$H$,\rmh  & 11.265$\pm$0.004 & 0.012 & 11.258$\pm$0.011 & 0.051 & 11.276$\pm$0.007 & 0.041 \\
$H$,\imh  & 11.247$\pm$0.005 & 0.017 & 11.240$\pm$0.011 & 0.053 & 11.260$\pm$0.008 & 0.045 \\
$K$,\bmk  & 11.283$\pm$0.009 & 0.029 & 11.272$\pm$0.008 & 0.044 & 11.284$\pm$0.007 & 0.044 \\
$K$,\vmk  & 11.277$\pm$0.009 & 0.029 & 11.269$\pm$0.008 & 0.045 & 11.280$\pm$0.007 & 0.045 \\
$K$,\rmk  & 11.280$\pm$0.009 & 0.029 & 11.278$\pm$0.008 & 0.045 & 11.288$\pm$0.007 & 0.044 \\
$K$,\imk  & 11.277$\pm$0.010 & 0.030 & 11.271$\pm$0.009 & 0.048 & 11.281$\pm$0.007 & 0.047 \\
$K$,\jmk  & 11.323$\pm$0.008 & 0.029 & 11.293$\pm$0.012 & 0.060 & 11.313$\pm$0.009 & 0.058 \\
\multicolumn{7}{c}{--- Three-bands ---} \\
$J$,\bmr  & 11.258$\pm$0.013 & 0.044 & 11.233$\pm$0.011 & 0.061 & 11.255$\pm$0.009 & 0.058 \\
$J$,\bmi  & 11.250$\pm$0.011 & 0.036 & 11.251$\pm$0.008 & 0.042 & 11.264$\pm$0.007 & 0.041 \\
$J$,\vmi  & 11.226$\pm$0.012 & 0.040 & 11.235$\pm$0.009 & 0.047 & 11.246$\pm$0.007 & 0.046 \\
$H$,\bmr  & 11.277$\pm$0.004 & 0.015 & 11.240$\pm$0.008 & 0.038 & 11.270$\pm$0.007 & 0.040 \\
$H$,\bmi  & 11.284$\pm$0.003 & 0.011 & 11.255$\pm$0.009 & 0.042 & 11.282$\pm$0.006 & 0.038 \\
$H$,\vmi  & 11.272$\pm$0.003 & 0.009 & 11.252$\pm$0.010 & 0.048 & 11.275$\pm$0.007 & 0.040 \\
$K$,\bmr  & 11.289$\pm$0.009 & 0.028 & 11.270$\pm$0.009 & 0.049 & 11.284$\pm$0.008 & 0.047 \\
$K$,\bmi  & 11.288$\pm$0.009 & 0.028 & 11.273$\pm$0.008 & 0.045 & 11.287$\pm$0.007 & 0.045 \\
$K$,\vmi  & 11.279$\pm$0.009 & 0.028 & 11.266$\pm$0.008 & 0.046 & 11.279$\pm$0.007 & 0.046 \\

\enddata
\tablenotetext{a}{Adopted PWZ relation.}
\tablenotetext{b}{Mean distance modulus and its error based on FO, FU and FU$+$FO variables.}
\tablenotetext{c}{Standard deviation of the distance modulus based on FO, FU and FU$+$FO 
variables.}
\label{tab:table_dmod_th_plw_nir}
\end{deluxetable}